\documentclass[twocolumn,superscriptaddress,prb]{revtex4-2}

\usepackage{dcolumn}
\usepackage{bm}
\usepackage{subfigure}
\usepackage[normalem]{ulem}
\usepackage{enumerate}
\usepackage{cancel}
\usepackage{flushend}
\usepackage{bbold}
\usepackage{mathtools}
\usepackage{float}
\usepackage{stmaryrd}
\usepackage{colortbl}
\usepackage[table]{xcolor}
\usepackage{array,ragged2e}
\usepackage{wasysym,graphicx,multirow,textcomp}
\usepackage{url}
\usepackage{romannum}
\usepackage{tabularx}
\usepackage{changes}
\usepackage[colorlinks = true,linkcolor = red,citecolor = magenta]{hyperref}
\usepackage[sort&compress]{natbib}

\usepackage{tikz-cd}

\newenvironment{helv}{\fontfamily{phv}\selectfont}{\par}
\DeclareTextFontCommand{\texthelv}{\helv}

\makeatletter
\newcommand{\thickhline}{%
    \noalign {\ifnum 0=`}\fi \hrule height 1pt
    \futurelet \reserved@a \@xhline
}
\newcolumntype{"}{@{\hskip\tabcolsep\vrule width 1pt\hskip\tabcolsep}}
\makeatother

\newcommand{\figsizeone}{0.8}
\newcommand{\figsizetwo}{1.0}

\begin{document}

\draft

\title{Biorthogonal scattering and generalized unitarity in non-Hermitian systems}

\author{Jung-Wan Ryu}
    \address{Center for Trapped Ion Quantum Science, Institute for Basic Science (IBS), Daejeon 34126, Republic of Korea}
    \address{Center for Theoretical Physics of Complex Systems, Institute for Basic Science (IBS), Daejeon 34126, Republic of Korea}    
    \address{Basic Science Program, Korea University of Science and Technology (UST), Daejeon 34113, Republic of Korea}

\author{Henning Schomerus}
    \address{Department of Physics, Lancaster University, Lancaster, LA1 4YB, United Kingdom}

\author{Hee Chul Park}
    \email{hc2725@gmail.com}
    \address{Department of Physics, Pukyong National University, Busan 48513, Republic of Korea}

\date{\today}

\begin{abstract}
We investigate the two-port scattering process in non-Hermitian dimer models via quantum measurements using external leads. We focus on two exemplary dimer models that preserve parity-time symmetry via spatial gain-loss balance and exhibit non-reciprocity due to directional hopping. The scattering matrix is constructed using the biorthogonality of the left and right scattering states of the Hamiltonian, allowing us to calculate the reflection and transmission probabilities. Our analysis compares the reflection and transmission coefficients derived from the left, right, and combined scattering states, revealing that, unlike in Hermitian systems, the non-Hermitian scattering process does not adhere to unitarity when considering only the right scattering states. Furthermore, non-Hermitian scattering can enhance the reflection and transmission probabilities, with distinct physical contributions arising independently from complex eigenvalues and the non-orthogonality of eigenstates. Our results clarify how biorthogonality restores generalized unitarity and identify distinct physical origins of enhanced transport in PT-symmetric and non-reciprocal dimers, providing new insights into quantum transport in non-Hermitian systems.
\end{abstract}

\maketitle

\section{Introduction}
\label{sec:intro}

Non-Hermitian systems have attracted significant interest due to their numerous potential applications and their ability to generalize the formalism of Hermitian quantum mechanics \cite{Bender1998, Mostafazadeh2002, Berry2004, Bender2007, Rotter2009, Moiseyev2011, ElGanainy2018, Ashida2020, Bergholtz2021}. Typically, the spectra of non-Hermitian systems are complex, the eigenstates are non-orthogonal, and the dynamics are non-unitary. These unique properties give rise to various exotic behaviors that do not appear in Hermitian systems \cite{Guo2009, Ruter2010, Lin2011, Liertzer2012, Brandstetter2014, Ramezani2014, Hodaei2017, Chen2017}. One particularly intriguing topic is quantum transport and wave scattering in non-Hermitian systems, as studied via measurements using external leads \cite{Schomerus_PRL2010_Quantum, Schomerus_PTRSA2013_From, PhysRevLett.105.053901, PhysRevLett.106.093902}. The resonant properties reflect the complex spectra and the intrinsic non-normality of non-Hermitian Hamiltonians, which can give rise to extraordinary phenomena such as directional and transient amplifications, respectively \cite{Makris2014, Nature2020, Xue2021, Makris2021, PhysRevA.108.052205}. Complex spectra imply the coexistence of amplifying and decaying modes, leading to directional amplification where the system preferentially enhances signals along a specific direction. In contrast, non-normality implies that even when all eigenmodes are decaying, constructive interference among non-orthogonal eigenstates can transiently amplify signals before eventual decay, 
a mechanism that underpins recent observations of non-Hermitian gain without conventional gain media.

Eigenstates of a non-Hermitian Hamiltonian \(H\) are generally non-orthogonal, where 
\(H|\psi_j \rangle = E_j | \psi_j \rangle\) and \(\langle \psi_j | \psi_i \rangle \ne \delta_{i,j}\), with \(\psi_j\) being the right eigenstates. Unlike the orthogonality of eigenstates in Hermitian Hamiltonians, non-Hermitian Hamiltonians exhibit biorthogonal eigenstates away from exceptional points, such that \(\langle \phi_j | \psi_i \rangle = \delta_{i,j}\), where \(\phi_j\) represents the left eigenstates, 
defined by 
\( \langle \phi_j | H = E_j \langle \phi_j | \) or 
\(H^{\dagger}|\phi_j \rangle = E^{*}_j | \phi_j \rangle\). 
Physical observables can be formulated in both the left-right (LR) and right-right (RR) bases. The RR basis is often used in practical calculations of observables, whereas the LR basis provides a mathematically consistent extension of orthogonality from Hermitian systems \cite{Mostafazadeh2010, Brody2013}. This distinction is crucial in non-Hermitian topology, where different basis choices may lead to distinct classifications and boundary phenomena \cite{Gong2018,Yao2018}. Recent works have explicitly highlighted that the LR and RR formulations do not necessarily yield equivalent physics: for example, in entanglement spectra of fermionic models, LR-based definitions faithfully inherit bulk topological invariants while RR-based ones can lead to distinct classifications \cite{Herviou2019}. Likewise, in quantum transport, LR and RR formalisms can predict opposite current-voltage responses, with only the RR framework yielding physically consistent results in PT-symmetric superconducting junctions \cite{Kornich2023}. Nevertheless, the precise physical interpretation and implications of choosing between the RR and LR bases in general non-Hermitian systems remain under active investigation.

In this work, we investigate the two-port scattering process in non-Hermitian dimer models through quantum measurement via external leads. The scattering matrix is constructed based on the biorthogonality of the left and right scattering states of the Hamiltonian. We examine the reflection and transmission coefficients derived from the left, right, and combined scattering states, respectively. Unlike Hermitian scattering processes, non-Hermitian scattering does not conform to unitarity when derived solely from the right scattering states \cite{Xu2023}. Instead, the non-Hermitian scattering process can lead to an enhancement of reflection and transmission probability, which represent physical quantities arising independently from the complex eigenvalues and non-orthogonal eigenstates of the Hamiltonian. These coefficients also mathematically satisfy the biorthogonal unitary process in non-Hermitian dimer models.

The structure of the paper is as follows. In Section~\ref{sec:QTNH}, we introduce two exemplary dimer models that preserve parity-time (PT) symmetry through spatial gain-loss balance and exhibit non-reciprocity (NR) due to directional hopping. By coupling the dimers with external leads, we construct the scattering matrix and calculate the reflection and transmission coefficients. In Section~\ref{sec:Results}, we present and discuss the results, emphasizing how they differ from scattering outcomes in Hermitian systems. Additionally, we compare the results in LR and RR bases. Finally, we summarize our findings in Section~\ref{sec:Summary}.

\section{Scattering in non-Hermitian systems}
\label{sec:QTNH}

\subsection{Scattering matrix}
\label{sec:Smatrix}

The scattering matrix plays a crucial role in describing how quantum or wave states evolve when particles or waves pass through a system. The scattering matrices for right and left scattering states read
\begin{eqnarray}
S^R = 
    \begin{pmatrix}
        r_{\rightarrow}^R & t_{\leftarrow}^R \\
        t_{\rightarrow}^R & r_{\leftarrow}^R
    \end{pmatrix}
    \mathrm{~and~}
S^L =
     \begin{pmatrix}
        r_{\rightarrow}^L & t_{\leftarrow}^L \\
        t_{\rightarrow}^L & r_{\leftarrow}^L
    \end{pmatrix}
    \label{smatrix} .
\end{eqnarray}
Here, the left and right scattering matrices are defined as linear maps between incoming and outgoing amplitudes in the asymptotic leads. For left scattering states, the incoming amplitudes are taken to multiply the scattering matrix from the right, consistent with the biorthogonal formulation. In the 1+1 channel geometry, the scattering matrix elements reduce to complex-valued reflection and transmission coefficients rather than matrices, where ${r_{\rightarrow(\leftarrow)}^{R(L)}}$ and ${t_{\rightarrow(\leftarrow)}^{R(L)}}$ are the reflection and transmission coefficients for right (left) scattering states when the incident waves come from the left (right) lead.

Since the $S^L$ is the same with $S^R$ in systems governed by Hermitian Hamiltonians, the scattering matrix is unitary, meaning it satisfies
\begin{equation}
{S^R}^{\dagger} S^R = I .
\end{equation}
where $S^{\dagger}$ is the Hermitian conjugate of the scattering matrix, and \( I \) is the identity matrix. That is,
\begin{eqnarray}
{r_{\rightarrow}^{R}}^{*} r_{\rightarrow}^R + {t_{\rightarrow}^{R}}^{*} t_{\rightarrow}^R &=& \bf{1} \\
{t_{\leftarrow}^{R}}^{*} t_{\leftarrow}^R + {r_{\leftarrow}^{R}}^{*} r_{\leftarrow}^R &=& \bf{1} \\
{r_{\rightarrow}^{R}}^{*} t_{\leftarrow}^R + {t_{\rightarrow}^{R}}^{*} r_{\leftarrow}^R &=& \bf{0} \\
{t_{\leftarrow}^{R}}^{*} r_{\rightarrow}^R + {r_{\leftarrow}^{R}}^{*} t_{\rightarrow}^R &=& \bf{0} .\end{eqnarray}
This unitarity ensures the conservation of probability, implying that no particle is lost during the scattering process. The sum of the transmission and reflection probabilities must always equal one, reflecting the fact that the total probability in a closed system without loss or gain is conserved.

This unitary property breaks down in non-Hermitian systems, where complex eigenvalues and non-orthogonal eigenstates can lead to non-conservation effects like gain or loss. As a result, the sum of transmission and reflection probabilities can deviate from 1, allowing for scenarios where the scattering process involves absorption or amplification. The breakdown of unitarity in the scattering matrix reflects the non-conservation of probability in non-Hermitian systems, leading to extraordinary behaviors like transient amplification or directional scattering.

Instead, in non-Hermitian cases, the scattering matrices satisfy
\begin{equation}
\label{Eq:non-Unitary}
{S^L}^{\dagger} S^R = I .
\end{equation}
For Hermitian leads, the generalized unitarity condition in Eq.~(\ref{Eq:non-Unitary}) follows directly from the orthogonality of asymptotic scattering states
defined in the leads.
That is,
\begin{eqnarray}
{r_{\rightarrow}^{L}}^{*} r_{\rightarrow}^R + {t_{\rightarrow}^{L}}^{*} t_{\rightarrow}^R &=& \bf{1} \\
{t_{\leftarrow}^{L}}^{*} t_{\leftarrow}^R + {r_{\leftarrow}^{L}}^{*} r_{\leftarrow}^R &=& \bf{1} \\
{r_{\rightarrow}^{L}}^{*} t_{\leftarrow}^R + {t_{\rightarrow}^{L}}^{*} r_{\leftarrow}^R &=& \bf{0} \\
{t_{\leftarrow}^{L}}^{*} r_{\rightarrow}^R + {r_{\leftarrow}^{L}}^{*} t_{\rightarrow}^R &=& \bf{0} .
\end{eqnarray}
where $S^R$ and $S^L$ are generally different.

\begin{figure}
    \centering
    \includegraphics[width=\figsizetwo\linewidth]{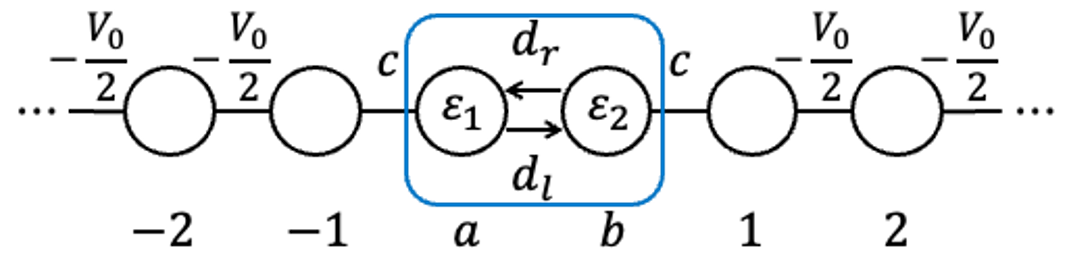}
    \caption{A dimer model with leads. Here, $a$ and $b$ represent the lattice labels for the system and positive and negative integer values denote the lattice labels for the right and left leads, respectively. Blue box represents a dimer model described by the system Hamiltonian $H_s$ of Eq.~(\ref{eq:Hamiltonian}).}
    \label{fig:QTNH_system}
\end{figure}

To better understand the above breakdown of unitarity, we briefly recall the general condition for unitary evolution in non-Hermitian systems, restricting our discussion to autonomous systems with no explicit time dependence in the Hamiltonian. Consider the expansion of a general state in the right-eigenbasis of the Hamiltonian,
\begin{equation}
|\psi(t)\rangle = \sum_n c_n e^{-i E_n t} |R_n\rangle ,
\end{equation}
where $H|R_n\rangle = E_n |R_n\rangle$. The norm then evolves as
\begin{equation}
\|\psi(t)\|^2 =
\sum_{m,n} c_m^* c_n \, e^{i(E_m^* - E_n)t}
\langle R_m | R_n \rangle .
\end{equation}
Requiring $\frac{d}{dt}\|\psi(t)\|^2=0$ for arbitrary initial states yields
\begin{equation}
\label{Eq:condition_uni}
(E_m^* - E_n)\,\langle R_m | R_n \rangle = 0
\qquad \forall \, m\neq n .
\end{equation}
Thus, unitary time evolution is recovered only when (i) all eigenvalues are real and (ii) the Hamiltonian is normal, meaning that its eigenstates form an orthogonal set. In other words, either complex eigenvalues or non-orthogonal eigenstates alone are sufficient to break unitarity. In non-Hermitian scattering, these two mechanisms correspond respectively to gain/loss and amplification originating from the overlap
of non-orthogonal modes associated with non-normal operators. The dimer models analyzed below illustrate these two distinct pathways to non-unitary behavior. This unified description clarifies that the biorthogonal scattering matrix framework of Eq.~(\ref{Eq:non-Unitary}) inherently captures both sources of non-unitarity in non-Hermitian systems. In the following subsections, we apply this formalism to concrete models, where the analytic structure of the scattering matrices, including poles and zeros, is analyzed explicitly.

\subsection{Hermitian-lead measurement and net flux imbalance}

We consider a two-terminal mesoscopic conductor connected to Hermitian semi-infinite leads. In this situation, the experimentally accessible quantities are defined operationally as ratios of conserved lead fluxes, probability currents, in the leads. The scattering states are expressed by 
\begin{equation}
\begin{pmatrix}
a_{I}^{out}(E)\\[2pt]
a_{II}^{out}(E)
\end{pmatrix}
=
\begin{pmatrix}
r_{\rightarrow}^R(E) & t_{\leftarrow}^R(E)\\
t_{\rightarrow}^R(E) & r_{\leftarrow}^R(E)
\end{pmatrix}
\begin{pmatrix}
a_{I}^{in}(E)\\[2pt]
a_{II}^{in}(E)
\end{pmatrix},
\label{eq:S_matrix_2terminal}
\end{equation}
whose scattering coefficients are functions of the incident energy ($E\in\mathbb{R}$). For a single propagating channel and an incoming state from the left, the asymptotic wave functions are
\begin{align}
\psi_{Ij} &\sim a_{I}^{in} e^{+ik_{I}j} + a_{I}^{out} e^{-ik_{I}j}, ~~ (j\to -\infty),\\
\psi_{IIj} &\sim a_{II}^{out} e^{+ik_{II}j}, ~~~~~~~~~~~~~~~~~ (j\to +\infty).
\end{align} 
The probability current is conserved in each Hermitian lead, which allows us to define measurable flux ratios, and is given by
\begin{equation}
J(E)=v(E)\big(|a^{\rm in}(E)|^2-|a^{\rm out}(E)|^2\big),
\label{eq:J_lead}
\end{equation}
where $v(E)$ is the group velocity at incident energy $E$, assuming it is identical in the two leads. This leads to the physically measurable reflectance and transmittance flux ratios
\begin{align}
R_{\rightarrow}^R(E) &\equiv \frac{J_{\rm ref}}{J_{\rm in}}
=\frac{|a_{I}^{\rm out}|^2}{|a_{I}^{\rm in}|^2}=|r_{\rightarrow}^R(E)|^2,
\label{eq:Rphys}\\
T_{\rightarrow}^R(E) &\equiv \frac{J_{\rm tr}}{J_{\rm in}}
=\frac{|a_{II}^{\rm out}|^2}{|a_{I}^{\rm in}|^2}
=|t_{\rightarrow}^R(E)|^2,
\label{eq:Tphys}
\end{align}
where $r_{\rightarrow}^R(E)\equiv a_I^{\rm out}/a_I^{\rm in}$ and $t_{\rightarrow}^R(E)\equiv a_{II}^{\rm out}/a_I^{\rm in}$ are the usual complex scattering amplitudes and we assume single-channel identical leads, $v_{I}=v_{II}$. In the same way, we can define the complex scattering amplitudes using an incoming state from right, $r_{\leftarrow}^R(E)\equiv a_{II}^{\rm out}/a_{II}^{\rm in}$ and $t_{\leftarrow}^R(E)\equiv a_{I}^{\rm out}/a_{II}^{\rm in}$. Equations~\eqref{eq:Rphys}--\eqref{eq:Tphys} are the mesoscopic analog of probability statements: $R$ and $T$ are nonnegative real numbers because they are ratios of lead currents, not bilinear forms in a non-Hermitian metric.

If the scatterer is Hermitian and time-independent, the $2\times2$ scattering matrix is unitary and lead flux is conserved, implying $R_{\rightarrow}^R+T_{\rightarrow}^R=1$. In contrast, when the effective scattering region is
non-Hermitian due to absorption, radiative loss, coupling to unobserved degrees of freedom, or active pumping, the two-terminal scattering matrix is generally non-unitary and the outgoing lead flux differs from the incoming one. We therefore introduce the net flux imbalance \cite{Schomerus_PTRSA2013_From}
\begin{equation}
A_{\rightarrow}^R(E)\equiv 1-\big(R_{\rightarrow}^R(E)+T_{\rightarrow}^R(E)\big).
\label{eq:Aprob}
\end{equation}
From the mesoscopic measurement perspective, the above equation has a direct probabilistic meaning: for an incoming flux normalized to unity, $R_{\rightarrow}^R$ and $T_{\rightarrow}^R$ quantify the fractions of the
incident particles that are detected in the reflected and transmitted lead channels, while $A_{\rightarrow}^R$ quantifies the remaining fraction that is not accounted for by the two measured channels.
Consequently,
\begin{align}
\begin{cases}
A_{\rightarrow}^R(E)>0 &~~~\Rightarrow~~~~ \text{net \emph{absorption} (loss),}\\
A_{\rightarrow}^R(E)<0 &~~~\Rightarrow~~~~ \text{net \emph{emission} (gain),}\\
A_{\rightarrow}^R(E)=0 &~~~\Rightarrow~~~~ \text{lead-flux conservation.}
\end{cases}
\end{align}
We emphasize that $A_{\rightarrow}^R$ is not an ad hoc correction; it is an operationally defined observable that quantifies the deviation from flux conservation in the measured Hermitian leads.

Equation~\eqref{eq:Aprob} can be understood by embedding the effective two-terminal problem into a larger unitary scattering description that includes additional unobserved channels representing the environment, radiation modes, auxiliary leads, or pump degrees of freedom. Denoting these channels collectively by $\nu$, one may write a total unitary scattering matrix $S_{\rm tot}$ acting on the direct sum of measured lead channels and unobserved channel. The total leakage probability into channels that are not monitored by the two-terminal measurement is following,
\begin{equation}
A_{\rightarrow}^R(E)=\sum_{\nu}\,T_{\nu\leftarrow I}(E),
\label{eq:A_as_leakage}
\end{equation}
where $T_{\nu\leftarrow I}$ is the flux fraction scattered from the left incident channel into the unobserved channels, $\nu$. 
In a purely absorbing (passive) environment all $T_{\nu\leftarrow I}\ge 0$ and hence $A_{\rightarrow}^R\ge 0$.
In the presence of pumping or gain, one may obtain $A_{\rightarrow}^R<0$, indicating net amplification in the measured channels.
A fully consistent quantum description then requires a Bogoliubov (pseudo-unitary) input--output formulation, in which annihilation and creation operators are mixed and the extended scattering matrix obeys $S^\dagger\Sigma_z S=\Sigma_z$ with $\Sigma_z=\mathrm{diag}(+\mathbb{I},-\mathbb{I})$ (annihilation/creation sectors) \cite{Beenakker1998}.

In a biorthogonal formulation of non-Hermitian scattering it is often convenient to introduce left and right scattering amplitudes and define bilinear quantities $R_{\rightarrow}^{bi}=r_{\rightarrow}^{L*} r_{\rightarrow}^R$ and $T_{\rightarrow}^{bi}=t_{\rightarrow}^{L*} t_{\rightarrow}^R$, which can satisfy an algebraic conservation law of the form $R_{\rightarrow}^{bi}+T_{\rightarrow}^{bi}=1$ under suitable generalized orthogonality conditions. These quantities, however, are not the flux ratios measured in Hermitian leads and can be complex. The experimentally relevant flux balance is instead governed by Eqs.~\eqref{eq:Rphys}--\eqref{eq:Aprob}, where $A$ provides the mesoscopic, probabilistic account of absorption ($A_{\rightarrow}^R>0$) or emission ($A_{\rightarrow}^R<0$) in the two-terminal measurement.

\subsection{Hamiltonian}

The total Hamiltonian of a dimer model with leads (Fig.~\ref{fig:QTNH_system}) consists of $H_s$, $H_l$, and $H_c$, which describe the system, the leads, and the coupling between them, respectively. The three contributions are as follows:
\begin{eqnarray}
H_s &=& \varepsilon_1 |a \rangle \langle a| + \varepsilon_2 |b \rangle \langle b| +  d_l |b \rangle \langle a| + d_r |a \rangle \langle b| \label{eq:Hamiltonian} \\
H_l &=& - \frac{V_0}{2}\Sigma_{j \ge 1} (|j \rangle \langle j+1| + |j+1 \rangle \langle j|) \\\nonumber
&& - \frac{V_0}{2}\Sigma_{j \le -1} (|j \rangle \langle j-1| + |j-1 \rangle \langle j|) \\
H_c &=&  c (|a \rangle \langle -1| + |b \rangle \langle 1| + |-1 \rangle \langle a| + |1 \rangle \langle b|),
\end{eqnarray}
where $\varepsilon_1$ and $\varepsilon_2$ are the complex onsite energies and $c$ is the coupling strength between the system and leads. In this work, $d_l$, $d_r$, $c$, and $V_0$ are real values.
The trial wave function is $| \psi \rangle = \phi_a |a \rangle + \phi_b |b \rangle + \Sigma_{j} \psi_j |j\rangle$, where $\phi_{\mu}$ denotes amplitudes on the dimer sites, $\mu=\{a, b\}$ and $\psi_j$ are wave amplitudes in the leads. The scattering states of the system are determined by the Schr\"odinger equation, $H |\psi \rangle = E |\psi \rangle $, where $|\psi \rangle$ is the right scattering state \cite{JWRyu_2017}.

\subsection{Reflection and transmission coefficients for right scattering states}

If the wavenumber in the leads corresponding to the incident energy $E$ is $q$, then the incident wavefunction from the left lead is given by 
\begin{eqnarray}
\psi_j &=& e^{i q j} + r_{\rightarrow}^R e^{- i q j},   ~~~ (j<0) \\
    &=& t_{\rightarrow}^R e^{i q j},  ~~~~~~~~~~~~~ (j>0)
\end{eqnarray}
where $r_{\rightarrow}^R$ and $t_{\rightarrow}^R$ are reflection and transmission coefficients according to the right eigenstates with incident wave from a left lead, respectively. From the dispersion relation of the lead, $e^{\pm i q} = -E /V_0\pm i \sqrt{1-|E/V_{0}|^2}$. The matrix equation of the right scattering coefficients is
\begin{eqnarray}
\label{matrixform_rl}
    \begin{pmatrix}
        V_0/2            & c_1^{T} & 0 \\
        c_1 e^{i q} & H_s-EI         & c_2 e^{i q} \\
        0                & c_2^{T}     & V_0/2
    \end{pmatrix}
    \begin{pmatrix}
        r_{\rightarrow}^R \\
        \Phi_{\rightarrow}^R   \\
        t_{\rightarrow}^R
    \end{pmatrix} 
    = 
    \begin{pmatrix}
        - V_0/2\\
        -c_1 e^{-i q}  \\
        0
    \end{pmatrix},
\end{eqnarray}
where $c_{1}=(c~~0)^{T}$, $c_{2}=(0~~c)^{T}$, $\Phi_{\rightarrow}^R=(\phi_{a \rightarrow}^R~~\phi_{b \rightarrow}^R)^T$. Simply, we can write the equation as $\bf{MX=B}$. Inverting the matrix $\bf{M}$, the reflection and transmission coefficients are
\begin{eqnarray}
r_{\rightarrow}^R &=& -\frac{(E-\varepsilon_1+2 \eta_{-})(E-\varepsilon_2+2 \eta_{+})-d_l d_r}{(E-\varepsilon_1+2 \eta_{+})(E-\varepsilon_2+2 \eta_{+})-d_ld_r}\\
t_{\rightarrow}^R &=& \frac{2d_l (\eta_{+} -\eta_{-})}{(E-\varepsilon_1+2 \eta_{+})(E-\varepsilon_2+2 \eta_{+})-d_ld_r},
\end{eqnarray}
where $\eta_{\pm}(E)\equiv (c^2/V_0) e^{\pm i q(E)}$ are self-energies from the coupling to the leads. The reflection probability $R_{\rightarrow}^R$ and transmission probability $T_{\rightarrow}^R$ are $|r_{\rightarrow}^R|^2$ and $|t_{\rightarrow}^R|^2$, respectively.

The incident wavefunction from the right lead is given by
\begin{eqnarray}
\psi_j &=& e^{-i q j} + r_{\leftarrow}^R e^{i q j},   ~~~ (j>0) \\
    &=& t_{\leftarrow}^R e^{-i q j},  ~~~~~~~~~~~~~ (j<0)
\end{eqnarray}
where $r_{\leftarrow}^R$ and $t_{\leftarrow}^R$ are reflection and transmission coefficients, respectively, and they are obtained as follows,
\begin{eqnarray}
r_{\leftarrow}^R &=& -\frac{(E-\varepsilon_1+2 \eta_{+})(E-\varepsilon_2+2 \eta_{-})-d_ld_r}{(E-\varepsilon_1+2 \eta_{+})(E-\varepsilon_2+2 \eta_{+})-d_ld_r}\\
t_{\leftarrow}^R &=& \frac{2d_r (\eta_{+} - \eta_{-})}{(E-\varepsilon_1+2 \eta_{+})(E-\varepsilon_2+2 \eta_{+})-d_l d_r}.
\end{eqnarray}

\subsection{Reflection and transmission coefficients for left scattering states}

Next, we consider left scattering states. When we consider the incident wavefunction from the left lead, the matrix equation of the left scattering coefficients is
\begin{eqnarray}
    \begin{pmatrix}
        V_0/2            & c_1^{T} & 0 \\
        c_1 e^{i q} & H_s^{\dagger}-E I         & c_2 e^{i q} \\
        0                & c_2^{T}     & V_0/2
    \end{pmatrix}
    \begin{pmatrix}
        r_{\rightarrow}^L \\
        \Phi_{\rightarrow}^L   \\
        t_{\rightarrow}^L
    \end{pmatrix} 
    = 
    \begin{pmatrix}
        - V_0/2\\
        -c_1 e^{-i q}  \\
        0
    \end{pmatrix}.
\end{eqnarray}
The reflection coefficient $r_{\rightarrow}^L$ and transmission coefficient $t_{\rightarrow}^L$ for the left scattering states are the same as those for the right scattering states when $\varepsilon_j \rightarrow \varepsilon_j^{*}$, $d_l \rightarrow d_r$, and $d_r \rightarrow d_l$.
Inverting the matrix $\bf{M}$, the reflection and transmission coefficients are 
\begin{eqnarray}
r_{\rightarrow}^L &=& -\frac{(E-\varepsilon_1^*+2 \eta_{-})(E-\varepsilon_2^*+2 \eta_{+})-d_l d_r}{(E-\varepsilon_1^*+2 \eta_{+})(E-\varepsilon_2^*+2 \eta_{+})-d_ld_r}\\
t_{\rightarrow}^L &=& \frac{2 d_r (\eta_{+}-\eta_{-})}{(E-\varepsilon_1^*+2 \eta_{+})(E-\varepsilon_2^*+2 \eta_{+})-d_l d_r}.
\end{eqnarray}

When we consider the incident wavefunction from the right lead, the reflection coefficient $r_{\leftarrow}^L$ and transmission coefficient $t_{\leftarrow}^L$ for the left scattering states are the same as those for the right scattering states when $\varepsilon_j \rightarrow \varepsilon_j^{*}$, $d_l \rightarrow d_r$, and $d_r \rightarrow d_l$ as follows,
\begin{eqnarray}
r_{\leftarrow}^L &=& -\frac{(E-\varepsilon_1^*+2 \eta_{+})(E-\varepsilon_2^*+2 \eta_{-})-d_ld_r}{(E-\varepsilon_1^*+2 \eta_{+})(E-\varepsilon_2^*+2 \eta_{+})-d_ld_r}\\
t_{\leftarrow}^L &=& \frac{2d_l (\eta_{+}-\eta_{-})}{(E-\varepsilon_1^*+2 \eta_{+})(E-\varepsilon_2^*+2 \eta_{+})-d_ld_r}.
\end{eqnarray}
One can directly verify that the scattering coefficients obtained above satisfy the biorthogonality condition ${S^L}^{\dagger} S^R = I$ in Eq.~(\ref{Eq:non-Unitary}) for this model.

\section{Results}
\label{sec:Results}

\subsection{A single site model}
\label{sec:single}

\begin{figure}
    \centering
    \includegraphics[width=\figsizeone\linewidth]{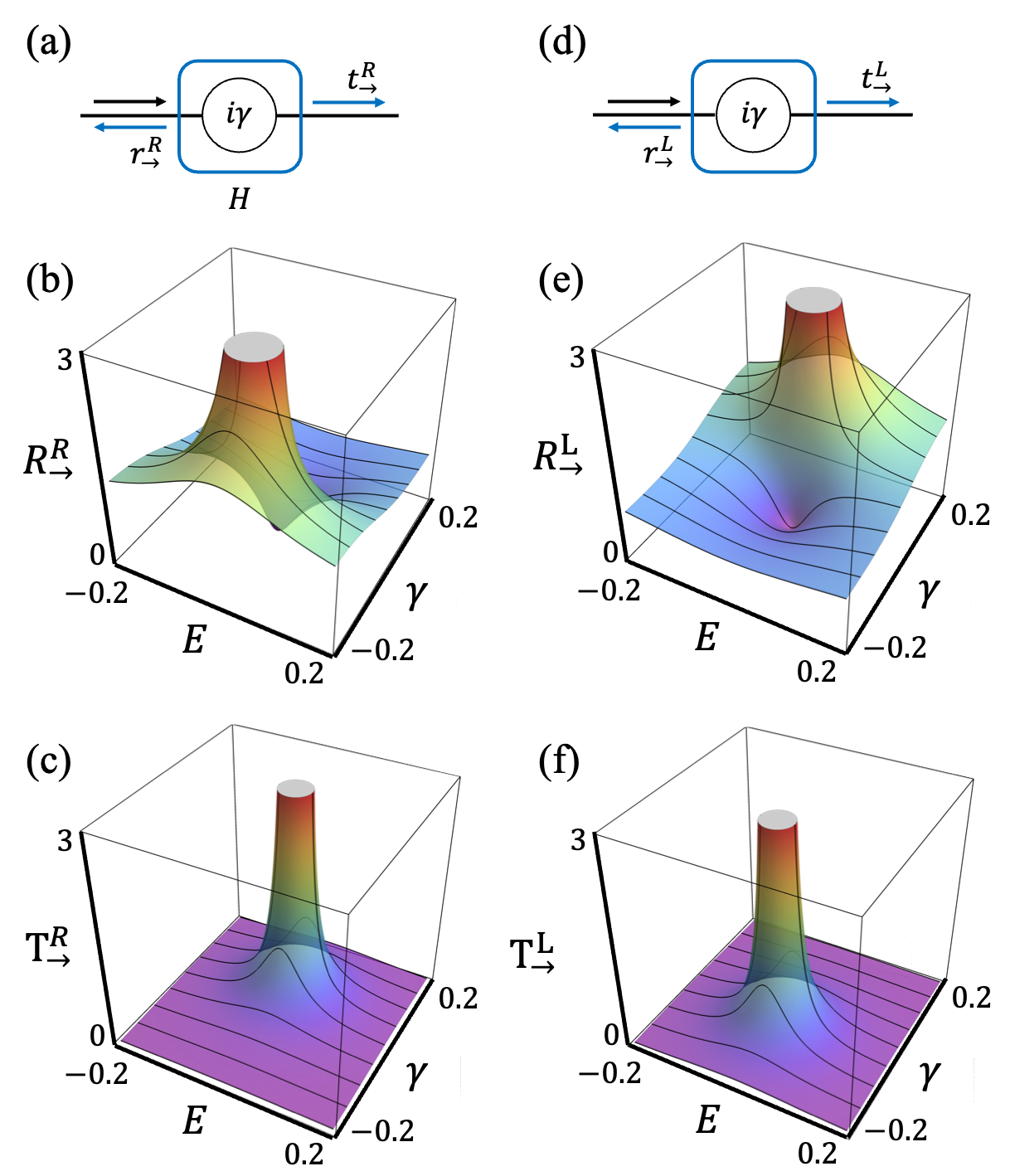}
    \caption{(a) is a schematic figure for the right scattering process. (b) and (c) are reflection and transmission probabilities for the right scattering states of the single-site model indicated in the top panels with incident wave from a left lead as a function of ($E$, $\gamma$), respectively. (d)-(f) are the case of the left scattering process. The results with incident waves from left and right leads are the same. The left scattering states of $H$ correspond to the right scattering states of $H^\dagger$.}
    \label{fig:fig_single}
\end{figure}

We first consider a single-site model (Fig.~\ref{fig:fig_single}) to reveal the effect of a non-Hermitian Hamiltonian with complex eigenenergies on the transmission and reflection probabilities. In Fig.~\ref{fig:fig_single}, the scattering energy \(E\) is kept real, while the imaginary part of the complex eigenenergy is controlled by the imaginary onsite term \(i \gamma\). The reflection probabilities for the right scattering states exhibit a peak at \(\gamma_0 = 2 c^2 / V_0 \) and a dip at 0, originating from the coupling between the system and the leads. The transmission probabilities also show a peak at \(\gamma_0\). Unlike in Hermitian systems, both probabilities can exceed 1 due to non-Hermiticity, which results in non-zero imaginary parts of the complex energies. As \(\gamma\) increases beyond \(\gamma_0\), the transmission probability decreases due to off-resonant conditions; however, in this regime the system becomes dynamically unstable, as the scattering matrix develops a pole in the upper half of the complex energy plane \cite{PhysRevB.54.11887}. The reflection and transmission probabilities for the left scattering states exhibit the same properties but with the opposite sign of \(\gamma\), corresponding to the onsite energy of \(H^+\) being \(-i \gamma\).

\subsection{Parity-time symmetric dimer}
\label{sec:PT dimer}

\begin{figure}
    \centering
    \includegraphics[width=\figsizeone\linewidth]{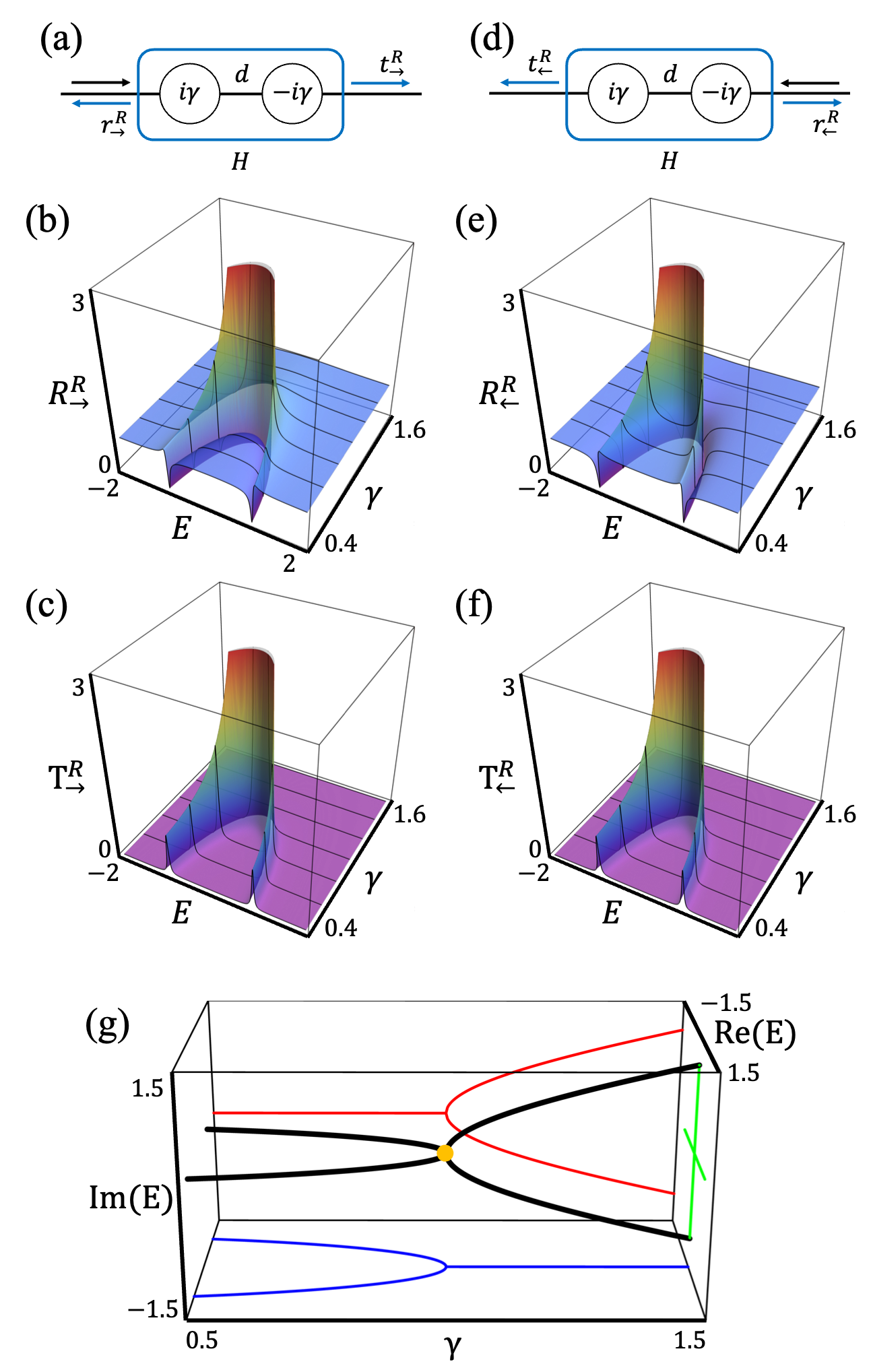}
    \caption{(a) is a schematic figure for the right scattering process with incident wave from left leads attached to a PT dimer. (b) and (c) are reflection and transmission probabilities for the right scattering states as a function of ($E$, $\gamma$), respectively. (d)-(f) are the case of the right scattering process with incident wave from right leads.
    There is an EP when $\gamma = 1$. (g) Evolution of two eigenenergies connected via EP in a PT dimer model without leads. As $\gamma$ increases, two real eigenenergies (solid black) approach each other and then coalesce at an EP (large orange dot). They split into two complex conjugate eigenenergies from the EP. The three projected figures show the real (blue) and imaginary (red) parts of the complex eigenenergies as a function of $\gamma$ and complex energy (green).}
    \label{fig:fig_PT_re}
\end{figure}

\begin{figure}
    \centering
    \includegraphics[width=\figsizeone\linewidth]{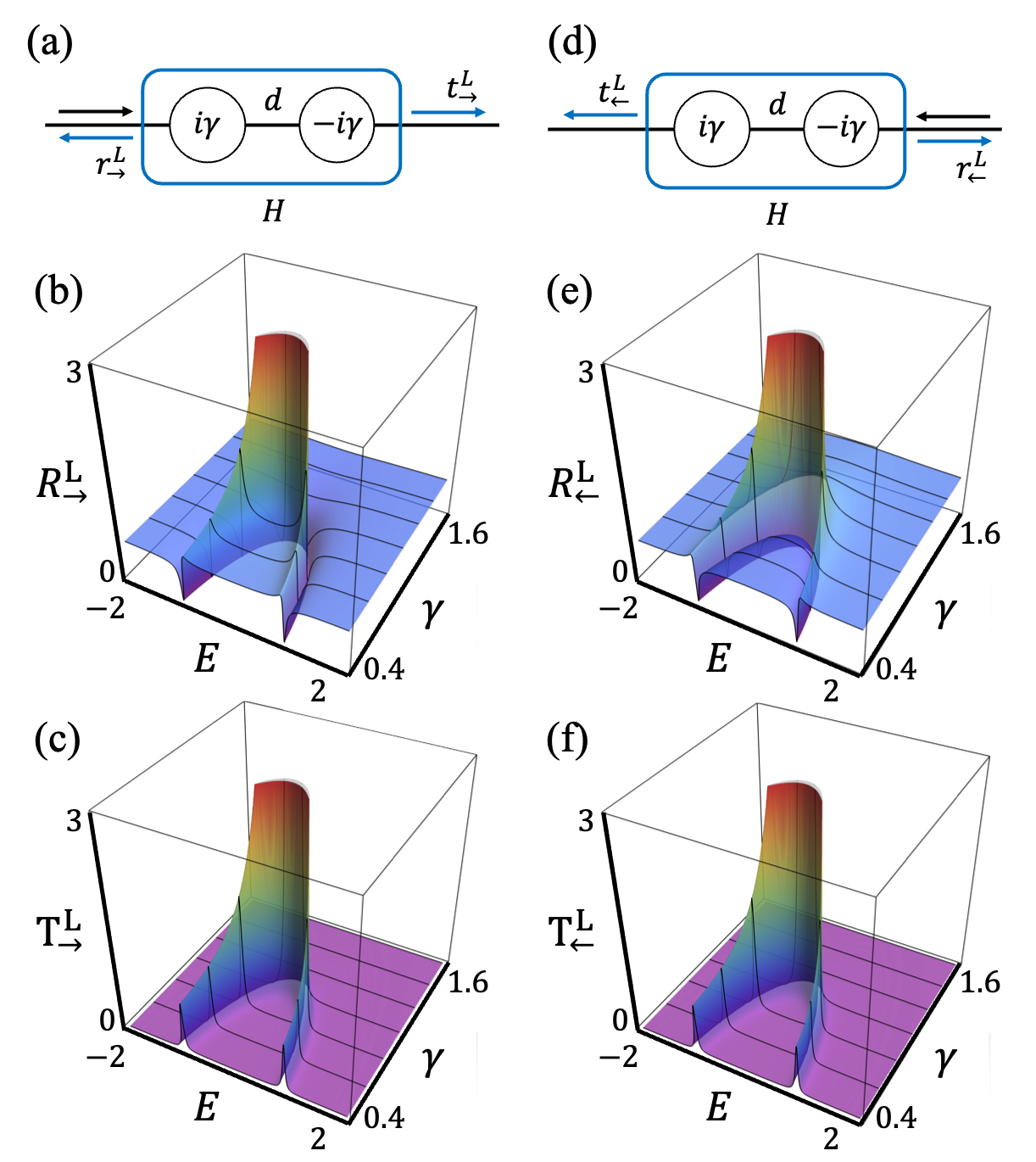}
    \caption{(a) is a schematic figure for the left scattering process with incident wave from left leads attached to a PT dimer. (b) and (c) are reflection and transmission probabilities for the left scattering states as a function of ($E$, $\gamma$), respectively. (d)-(f) are the case of the left scattering process with incident wave from right leads. Results are the same as those in Fig.~\ref{fig:fig_PT_re} with opposite incident waves.}
    \label{fig:fig_PT_le}
\end{figure}

\begin{figure}
    \centering
    \includegraphics[width=\figsizetwo\linewidth]{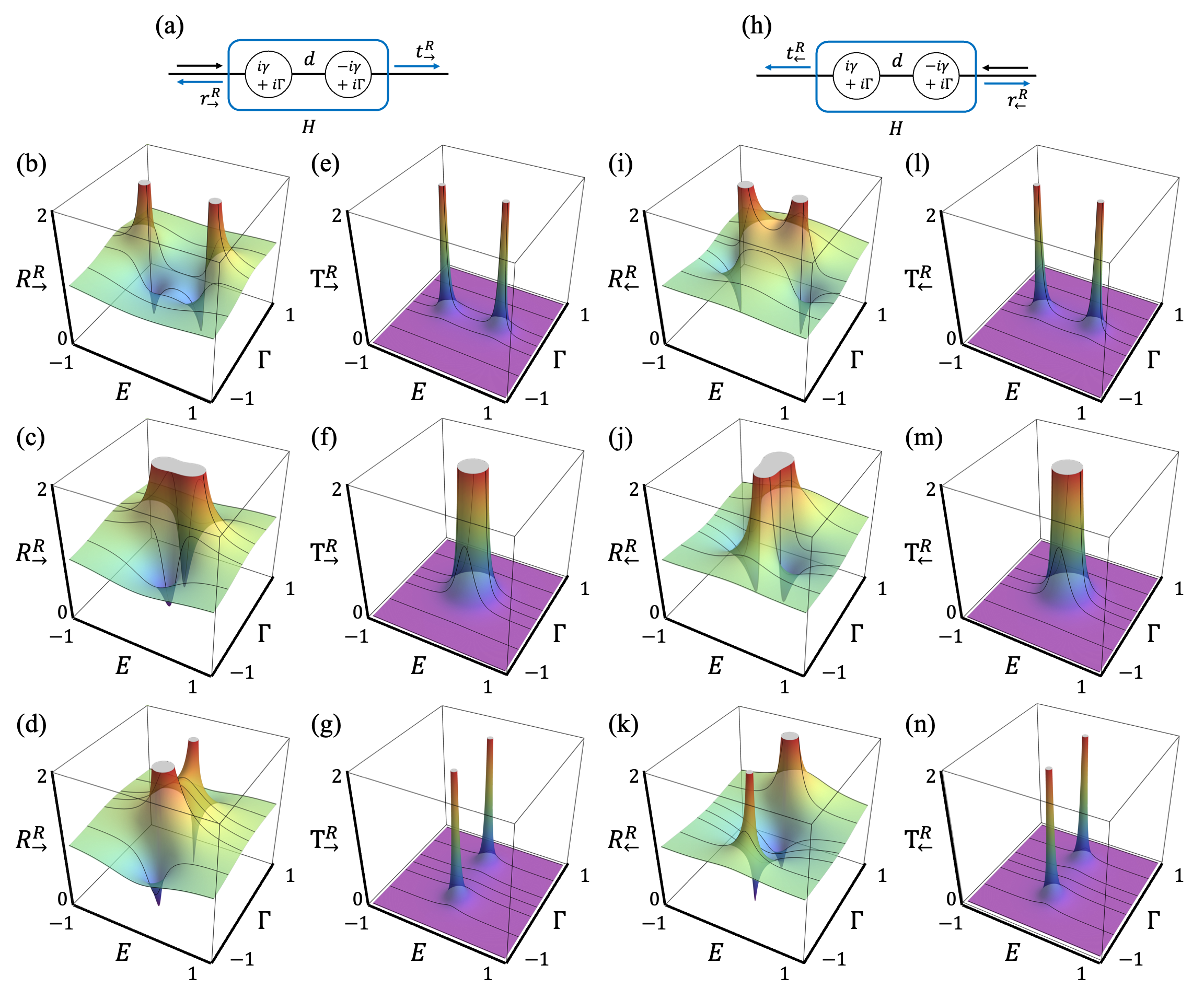}
    \caption{(a) is a schematic figure for the right scattering process with incident wave from left leads attached to a PT dimer with additional overall loss or gain $\Gamma$. (b)-(d) are reflection probabilities for the right scattering states as a function of ($E$, $\Gamma$) for $\gamma=\{0.9, 1.0,\text{ and }1.1\}$. (e)-(g) are transmission probabilities. (h)-(n) are the case of the right scattering process with incident wave from right leads.
    In case of opposite incident waves, reflection probabilities are different but transmission probabilities are the same. There is an EP when $\gamma = 1$.}
    \label{fig:fig_PT_Gamma}
\end{figure}

We now consider a PT-dimer (Fig.~\ref{fig:fig_PT_re}) for which Hamiltonian is given in Eq.~(\ref{eq:Hamiltonian}) with $d_r = d_l = d$ and $\varepsilon_1 = -\varepsilon_2 = i \gamma$ 
\begin{eqnarray}
H_s=
\begin{pmatrix}
        i\gamma  &          d \\
        d        & -i\gamma       
    \end{pmatrix}.
\end{eqnarray}
As $\gamma$ increases in the system Hamiltonian $H_s$ while preserving $\mathcal{PT}$ symmetry, two real eigenvalues of $H_s$ approach each other for $\gamma < 1$, merge at an exceptional point (EP) of $H_s$ at $\gamma = 1$, where the corresponding eigenstates also coalesce, and subsequently split into two purely imaginary eigenvalues for $\gamma > 1$. Reflection and transmission probabilities for right (left) scattering state exhibit Fano-type and resonant peaks at real eigenenergies, respectively (Fig.~\ref{fig:fig_PT_re} and Fig.~\ref{fig:fig_PT_le}). Reflection probabilities for right scattering states with an incident wave from the left (right) lead correspond to those for left scattering states with an incident wave from the right (left) lead, while all transmission probabilities are the same independent of left and right scattering states and directions of incident waves.

To clarify the underlying pole and zero structure of the PT dimer, we explicitly evaluate the denominator and numerator of reflection and transmission coefficients. The common denominator becomes
\[
D(E, \gamma) = (E - i\gamma + 2\eta_+)(E + i\gamma + 2\eta_+) - d^2,
\]
which gives
\[
(E + 2\eta_+)^2 + \gamma^2 = d^2.
\]
This defines an \emph{implicit relation} between $E$ and $\gamma$, forming a \emph{quasi-elliptic trajectory} in the $(E,\gamma)$ plane, as shown in Figs.~\ref{fig:fig_PT_re} and~\ref{fig:fig_PT_le}. 
The numerator for the reflection coefficient,
\[
N(E, \gamma) = (E - i\gamma + 2\eta_{\mp})(E + i\gamma + 2\eta_{\pm}) - d^2,
\]
with $\eta_- = \eta_+^*$, yields
\[
(E + 2\mathrm{Re}(\eta_{\pm}))^2 + (\gamma + 2\mathrm{Im}(\eta_{\pm}))^2 = d^2,
\]
indicating that reflection zeros lie on a shifted implicit curve. 
Therefore, poles and zeros appear at different complex positions, consistent with the separated loci of resonant and antiresonant responses in Figs.~\ref{fig:fig_PT_re} and \ref{fig:fig_PT_le}.

Assuming $E$ is a complex value, $\varepsilon + i \Gamma$, where $\Gamma$ can be considered overall gain or loss of the systems, we can find complex resonant peaks of reflection and transmission probabilities in the complex energy plane (Fig.~\ref{fig:fig_PT_Gamma}). The eigenenergies are shifted in the direction of imaginary energy by the amount of overall gain or loss $\Gamma$ from the eigenenergies of the PT dimer. By changing $\Gamma$, the reflection and transmission probabilities can be measured as a function of $\varepsilon$ and $\Gamma$, with results consistent with the controlling of complex incident energy in the PT dimer. Consequently, the measurement of scattering with $\Gamma$ control in the system plays the role of measurement with controlling gain and loss in both leads. This mapping between the control of $\Gamma$ and the tuning of complex incident energy explicitly clarifies how external gain or loss in the leads corresponds to complex energy modulation in the PT dimer.

\subsection{Non-reciprocal dimer}
\label{sec:NR dimer}

\begin{figure}
    \centering
    \includegraphics[width=\figsizeone\linewidth]{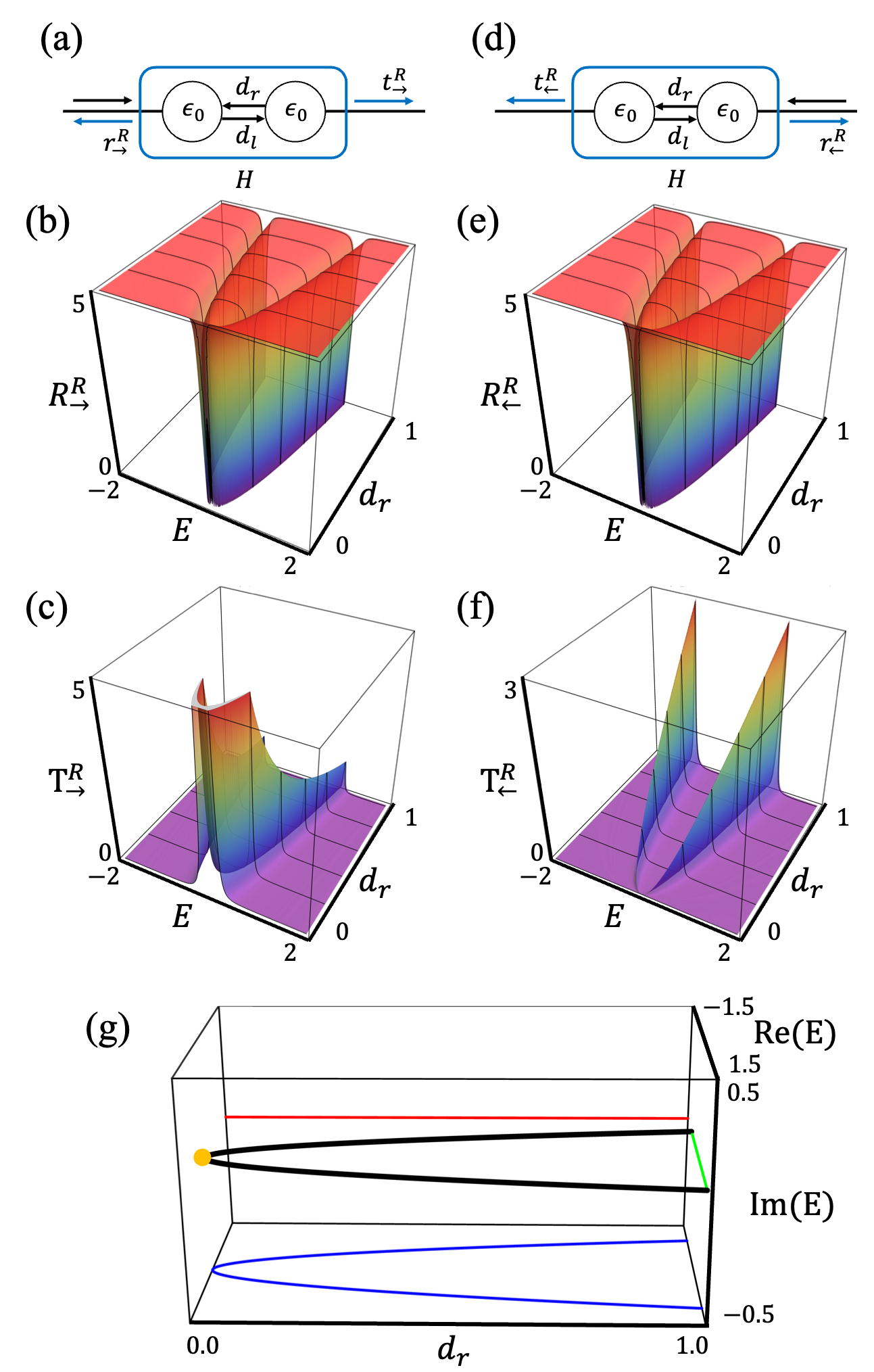}
    \caption{(a) is a schematic figure for the right scattering process with incident wave from left leads attached to a non-reciprocal dimer. (b) and (c) are reflection and transmission probabilities for the right scattering states as a function of ($E$, $d_r$), respectively. (d)-(f) are the case of the right scattering process with incident wave from right leads.
    There is an EP when $d_r = 0$. (g) Evolution of two eigenenergies connected via EP in a non-reciprocal dimer model without leads. As $d_r$ decreases, two real eigenvalues (solid black) approach each other and then coalesce at an EP (large orange dot) when $d_r = 0$. The three projected figures show the real (blue) and imaginary (red) parts of the complex eigenenergies as a function of $d_r$ and complex energy (green).}
    \label{fig:fig_NR_re}
\end{figure}

\begin{figure}
    \centering
    \includegraphics[width=\figsizeone\linewidth]{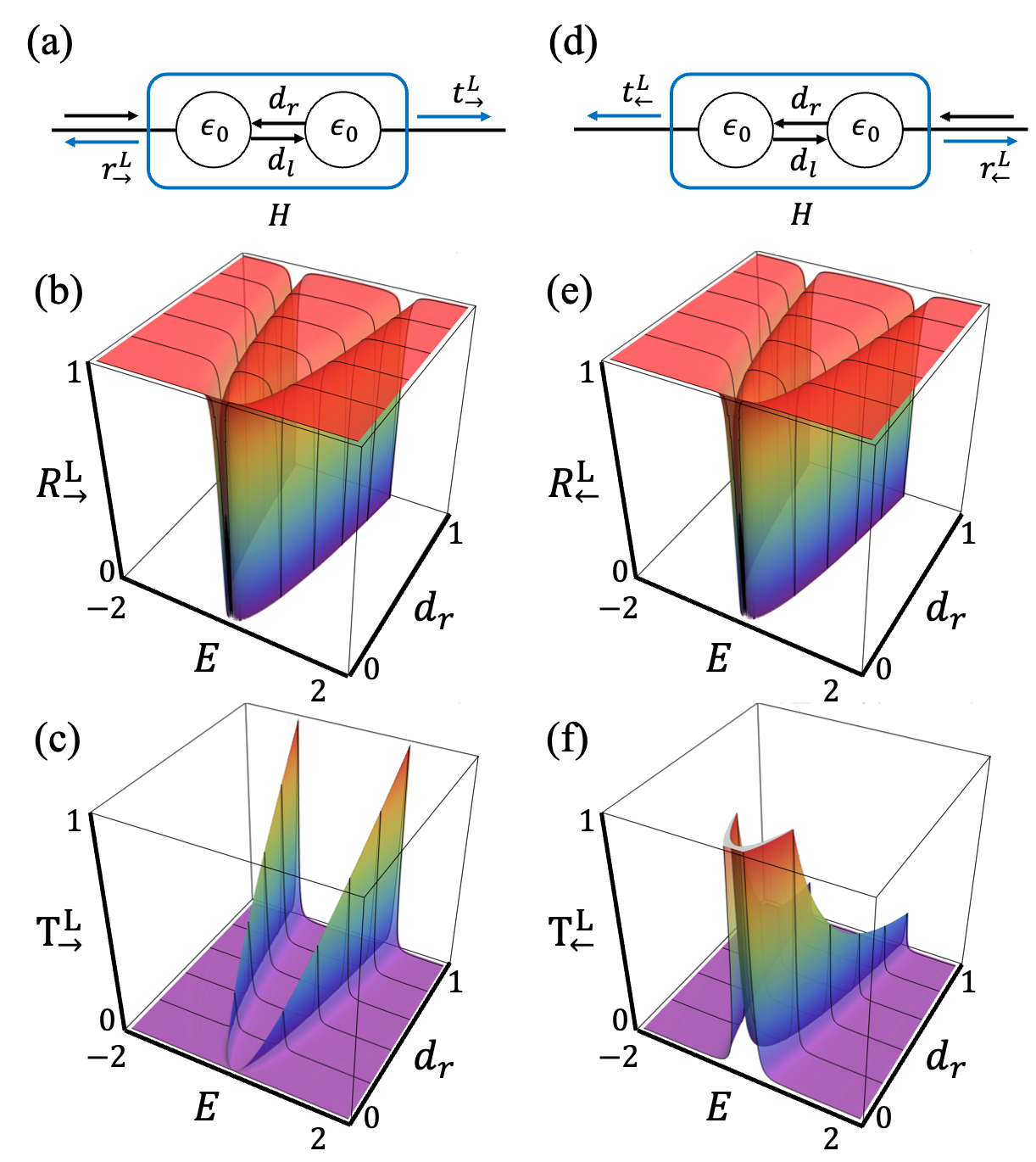}
    \caption{(a) is a schematic figure for the left scattering process with incident wave from left leads attached to a non-reciprocal dimer. (b) and (c) are reflection and transmission probabilities for the left scattering states as a function of ($E$, $d_r$), respectively. (d)-(f) are the case of the left scattering process with incident wave from right leads. Results are the same as those in Fig.~\ref{fig:fig_NR_re} with opposite incident waves.}
    \label{fig:fig_NR_le}
\end{figure}

\begin{figure}
    \centering
    \includegraphics[width=\figsizeone\linewidth]{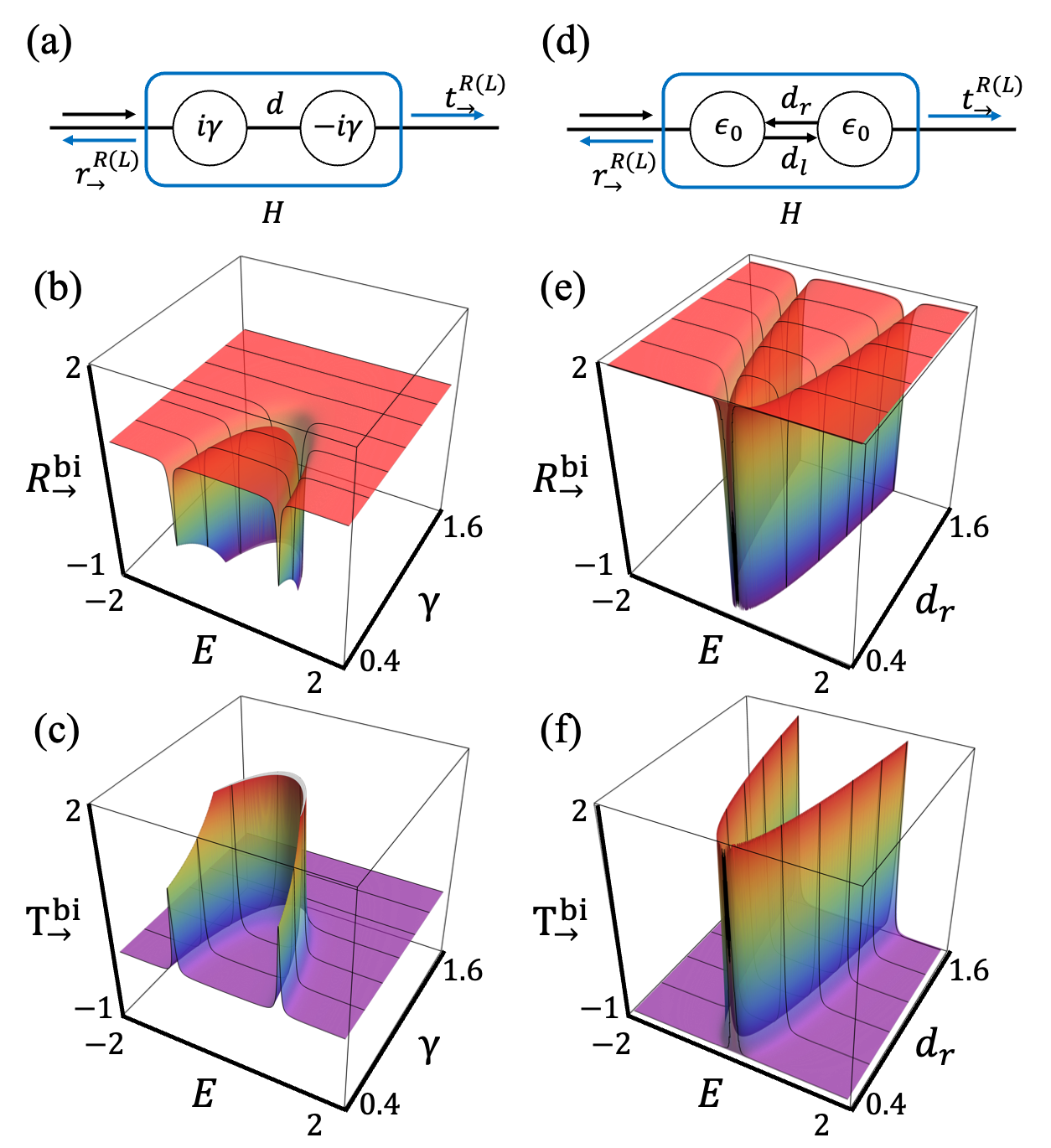}
    \caption{(a) is a schematic figure for the right/left scattering process with incident wave from a left lead. (b) and (c) are reflection and transmission probabilities for the biorthogonal scattering states with a PT dimer as a function of ($E$, $\gamma$), respectively. (d)-(f) are the case of a non-reciprocal dimer. The generalized unitarity condition $R_{\rightarrow}^{bi} + T_{\rightarrow}^{bi} = 1$ is always satisfied.}
    \label{fig:fig_PT_NR_bi}
\end{figure}

Next we consider a NR-dimer (Fig.~\ref{fig:fig_NR_re}) for which Hamiltonian is given in Eq.~(\ref{eq:Hamiltonian}) with $d_l = 1$ and $\varepsilon_1 = \varepsilon_2 = 0$. 
\begin{eqnarray}
H_s=
\begin{pmatrix}
        0  &       d_r  \\
        d_l        & 0       
    \end{pmatrix}.
\end{eqnarray}
As $d_r$ decreases from $1$, two real eigenvalues approach each other ($d_r > 0$) and finally they merge at an EP ($d_r = 0$). Reflection probabilities for both right and left scattering states exhibit dips at real eigenenergies, while transmission probabilities display resonant peaks (Fig.~\ref{fig:fig_NR_re} and Fig.~\ref{fig:fig_NR_le}). All reflection probabilities are the same independent of left and right scattering states and directions of incident waves but transmission probabilities for right scattering states with incident wave from left (right) lead correspond to those for left scattering states with incident wave from right (left) lead.

To elucidate the pole-zero structure of the non-reciprocal dimer, we evaluate the denominator and numerator of the scattering coefficients. 
The common denominator yields the pole condition
\[
D(E,d_r)=(E+2\eta_+)^2 - d_r = 0,
\]
while the reflection numerator gives the zero condition
\[
N(E,d_r)=(E+2\eta_{\mp})(E+2\eta_{\pm}) - d_r = 0.
\]
In the real-energy plots of Figs.~\ref{fig:fig_NR_re} and~\ref{fig:fig_NR_le}, one has $\eta_-=\eta_+^*$, so the zero locus can be written as
\[
|E+2\eta_+|^2 = d_r.
\]
Under the assumption $V_0 \gg |E|$ and $V_0 \gg c$, the imaginary part of $E+2\eta_+(E)$ is parametrically small, implying
\[
|E+2\eta_+|^2 \approx (E+2\eta_+)^2.
\]
Consequently, the pole and zero loci,
\[
(E+2\eta_{+})^2 = d_r
\quad \text{and} \quad
|E+2\eta_+|^2 = d_r,
\]
appear nearly identical in the non-reciprocal case.

It should be noted that the large enhancement of the transmission probabilities as $d_r$ decreases originates entirely from the non-orthogonality of eigenstates rather than from the imaginary parts of the complex eigenvalues since the NR dimer Hamiltonian has real eigenvalues. By contrast, in a single site model (Sec.~\ref{sec:single}), the enhancement of transmission probability beyond $1$ results from non-zero imaginary parts of complex eigenvalues.

Figure~\ref{fig:fig_PT_NR_bi} illustrates the generalized unitarity relation for both PT-symmetric and non-reciprocal dimers. To confirm the validity of Eq.~(\ref{Eq:non-Unitary}), we obtain the biorthogonal reflection and transmission probabilities, $R_{\rightarrow}^{bi}$, $R_{\leftarrow}^{bi}$, $T_{\rightarrow}^{bi}$, and $T_{\leftarrow}^{bi}$. In all cases, the sum of reflection and transmission probabilities becomes $1$ because of the biorthogonal unitary conditions. The reflection probability can be negative in the PT dimer, whereas in the NR dimers, all probabilities remain between $0$ and $1$ (Fig.~\ref{fig:fig_PT_NR_bi}). In other words, the generalized unitarity corresponds to the conservation of biorthogonal flux, where the total flux defined in the LR framework remains constant even though individual right- or left-state probabilities are not conserved.

\subsection{Vertically arranged parity-time symmetric dimer}
\label{sec:PT dimer II}

\begin{figure}
    \centering
    \includegraphics[width=\figsizetwo\linewidth]{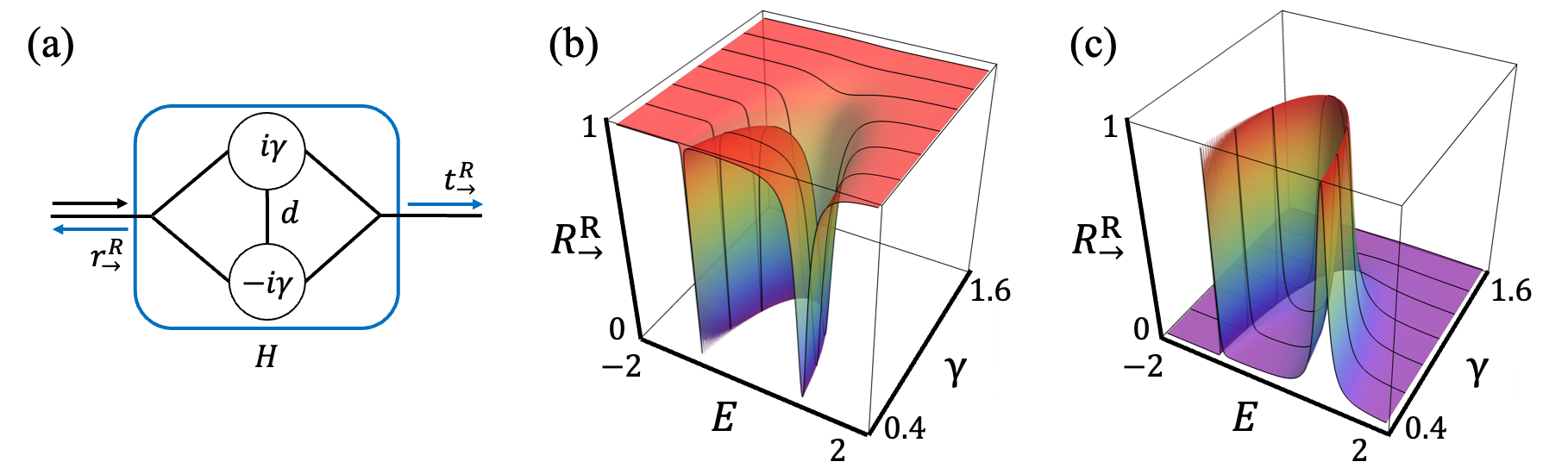}
    \caption{(a) is a schematic figure for the right scattering process with incident wave from a left lead attached to a vertical PT dimer case. (b) and (c) are reflection and transmission probabilities for right scattering states, respectively. The unitary condition $R_{\rightarrow}^{R} + T_{\rightarrow}^{R} = 1$ is satisfied due to the PT symmetry which is not broken by attached leads.}
    \label{fig:fig_vPT_re}
\end{figure}

In case of the vertically arranged PT dimer (Fig.~\ref{fig:fig_vPT_re}), we use the matrix equation Eq. (\ref{matrixform_rl}) to calculate the right scattering coefficients with $c_{1}=c_{2}=(c~~c)^{T}$ and the PT-symmetric system Hamiltonian as follows,
\begin{eqnarray}
H_s = 
    \begin{pmatrix}
        i\gamma  & d  \\
        d  & -i\gamma \\
    \end{pmatrix}.
\end{eqnarray}
The reflection probability $R$ and transmission probability $T$ are $|r_{\rightarrow}^R|^2$ and $|t_{\rightarrow}^R|^2$, respectively. In contrast to the previous PT dimer model, the sum of reflection and transmission probabilities in the RR basis is always $1$ since the coupling between system and leads does not break the PT-symmetry of the total Hamiltonian. When $\gamma < 1$, there are two resonant peaks corresponding to the real parts of complex energies (Fig.~\ref{fig:fig_vPT_re}), whose widths are very different because of the symmetry of the states about x-axis. The sharp and broad resonant peaks represent odd and even parity states, respectively.

\section{Summary}
\label{sec:Summary}

We have explored the two-port scattering process in non-Hermitian dimer models, focusing on those that preserve parity-time symmetry through spatial gain-loss balance and exhibit non-reciprocity due to directional hopping. By coupling the dimers to external leads, the scattering matrix is constructed using the biorthogonality of the left and right scattering states of the Hamiltonian, allowing the calculation of reflection and transmission coefficients based on the left, right, and combined scattering states.

The findings illustrate how, unlike the unitary scattering processes typical of Hermitian systems, non-Hermitian scattering processes do not adhere to unitarity when derived solely from the right scattering states. Additionally, the non-Hermitian scattering process can enhance reflection and transmission probabilities, which arise independently from the complex eigenvalues and non-orthogonal eigenstates of the Hamiltonian. These probabilities satisfy the biorthogonal unitary process in non-Hermitian dimer models, providing valuable insights into the unique behaviors of quantum transport and wave scattering in non-Hermitian systems.

\section*{Acknowledgments}
The authors acknowledge financial support from the Institute for Basic Science in the Republic of Korea through the project IBS-R024-D1. 
This work was supported from the National Research Foundation of Korea (NRF) Grant funded by the Ministry of Education (No. RS-2023-00285353, RS-2023-00256050) and Ministry of Science and ICT (No. RS-2025-00557045, RS-2023-00278511, RS-2025-02315685).

\bibliography{reference}

\begin{thebibliography}{36}%
\makeatletter
\providecommand \@ifxundefined [1]{%
 \@ifx{#1\undefined}
}%
\providecommand \@ifnum [1]{%
 \ifnum #1\expandafter \@firstoftwo
 \else \expandafter \@secondoftwo
 \fi
}%
\providecommand \@ifx [1]{%
 \ifx #1\expandafter \@firstoftwo
 \else \expandafter \@secondoftwo
 \fi
}%
\providecommand \natexlab [1]{#1}%
\providecommand \enquote  [1]{``#1''}%
\providecommand \bibnamefont  [1]{#1}%
\providecommand \bibfnamefont [1]{#1}%
\providecommand \citenamefont [1]{#1}%
\providecommand \href@noop [0]{\@secondoftwo}%
\providecommand \href [0]{\begingroup \@sanitize@url \@href}%
\providecommand \@href[1]{\@@startlink{#1}\@@href}%
\providecommand \@@href[1]{\endgroup#1\@@endlink}%
\providecommand \@sanitize@url [0]{\catcode `\\12\catcode `\$12\catcode `\&12\catcode `\#12\catcode `\^12\catcode `\_12\catcode `\%12\relax}%
\providecommand \@@startlink[1]{}%
\providecommand \@@endlink[0]{}%
\providecommand \url  [0]{\begingroup\@sanitize@url \@url }%
\providecommand \@url [1]{\endgroup\@href {#1}{\urlprefix }}%
\providecommand \urlprefix  [0]{URL }%
\providecommand \Eprint [0]{\href }%
\providecommand \doibase [0]{https://doi.org/}%
\providecommand \selectlanguage [0]{\@gobble}%
\providecommand \bibinfo  [0]{\@secondoftwo}%
\providecommand \bibfield  [0]{\@secondoftwo}%
\providecommand \translation [1]{[#1]}%
\providecommand \BibitemOpen [0]{}%
\providecommand \bibitemStop [0]{}%
\providecommand \bibitemNoStop [0]{.\EOS\space}%
\providecommand \EOS [0]{\spacefactor3000\relax}%
\providecommand \BibitemShut  [1]{\csname bibitem#1\endcsname}%
\let\auto@bib@innerbib\@empty
\bibitem [{\citenamefont {Bender}\ and\ \citenamefont {Boettcher}(1998)}]{Bender1998}%
  \BibitemOpen
  \bibfield  {author} {\bibinfo {author} {\bibfnamefont {C.~M.}\ \bibnamefont {Bender}}\ and\ \bibinfo {author} {\bibfnamefont {S.}~\bibnamefont {Boettcher}},\ }\bibfield  {title} {\bibinfo {title} {Real spectra in non-hermitian hamiltonians having pt symmetry},\ }\href {https://doi.org/10.1103/PhysRevLett.80.5243} {\bibfield  {journal} {\bibinfo  {journal} {Phys. Rev. Lett.}\ }\textbf {\bibinfo {volume} {80}},\ \bibinfo {pages} {5243} (\bibinfo {year} {1998})}\BibitemShut {NoStop}%
\bibitem [{\citenamefont {Mostafazadeh}(2002)}]{Mostafazadeh2002}%
  \BibitemOpen
  \bibfield  {author} {\bibinfo {author} {\bibfnamefont {A.}~\bibnamefont {Mostafazadeh}},\ }\bibfield  {title} {\bibinfo {title} {Pseudo-hermiticity versus pt symmetry: The necessary condition for the reality of the spectrum of a non-hermitian hamiltonian},\ }\href {https://doi.org/10.1063/1.1418246} {\bibfield  {journal} {\bibinfo  {journal} {Journal of Mathematical Physics}\ }\textbf {\bibinfo {volume} {43}},\ \bibinfo {pages} {205} (\bibinfo {year} {2002})}\BibitemShut {NoStop}%
\bibitem [{\citenamefont {Berry}(2004)}]{Berry2004}%
  \BibitemOpen
  \bibfield  {author} {\bibinfo {author} {\bibfnamefont {M.~V.}\ \bibnamefont {Berry}},\ }\bibfield  {title} {\bibinfo {title} {Physics of non-hermitian degeneracies},\ }\href {https://doi.org/10.1023/B:CJOP.0000044002.05657.04} {\bibfield  {journal} {\bibinfo  {journal} {Czech. J. Phys.}\ }\textbf {\bibinfo {volume} {54}},\ \bibinfo {pages} {1039} (\bibinfo {year} {2004})}\BibitemShut {NoStop}%
\bibitem [{\citenamefont {Bender}(2007)}]{Bender2007}%
  \BibitemOpen
  \bibfield  {author} {\bibinfo {author} {\bibfnamefont {C.~M.}\ \bibnamefont {Bender}},\ }\bibfield  {title} {\bibinfo {title} {Making sense of non-hermitian hamiltonians},\ }\href {https://doi.org/10.1088/0034-4885/70/6/R03} {\bibfield  {journal} {\bibinfo  {journal} {Rep. Prog. Phys.}\ }\textbf {\bibinfo {volume} {70}},\ \bibinfo {pages} {947} (\bibinfo {year} {2007})}\BibitemShut {NoStop}%
\bibitem [{\citenamefont {Rotter}(2009)}]{Rotter2009}%
  \BibitemOpen
  \bibfield  {author} {\bibinfo {author} {\bibfnamefont {I.}~\bibnamefont {Rotter}},\ }\bibfield  {title} {\bibinfo {title} {A non-hermitian hamilton operator and the physics of open quantum systems},\ }\href {https://doi.org/10.1088/1751-8113/42/15/153001} {\bibfield  {journal} {\bibinfo  {journal} {J. Phys. A: Math. Theor.}\ }\textbf {\bibinfo {volume} {42}},\ \bibinfo {pages} {153001} (\bibinfo {year} {2009})}\BibitemShut {NoStop}%
\bibitem [{\citenamefont {Moiseyev}(2011)}]{Moiseyev2011}%
  \BibitemOpen
  \bibfield  {author} {\bibinfo {author} {\bibfnamefont {N.}~\bibnamefont {Moiseyev}},\ }\href@noop {} {\emph {\bibinfo {title} {Non-Hermitian Quantum Mechanics}}}\ (\bibinfo  {publisher} {Cambridge University Press},\ \bibinfo {address} {Cambridge, England},\ \bibinfo {year} {2011})\BibitemShut {NoStop}%
\bibitem [{\citenamefont {El-Ganainy}\ \emph {et~al.}(2018)\citenamefont {El-Ganainy}, \citenamefont {Makris}, \citenamefont {Khajavikhan}, \citenamefont {Musslimani}, \citenamefont {Rotter},\ and\ \citenamefont {Christodoulides}}]{ElGanainy2018}%
  \BibitemOpen
  \bibfield  {author} {\bibinfo {author} {\bibfnamefont {R.}~\bibnamefont {El-Ganainy}}, \bibinfo {author} {\bibfnamefont {K.~G.}\ \bibnamefont {Makris}}, \bibinfo {author} {\bibfnamefont {M.}~\bibnamefont {Khajavikhan}}, \bibinfo {author} {\bibfnamefont {Z.~H.}\ \bibnamefont {Musslimani}}, \bibinfo {author} {\bibfnamefont {S.}~\bibnamefont {Rotter}},\ and\ \bibinfo {author} {\bibfnamefont {D.~N.}\ \bibnamefont {Christodoulides}},\ }\bibfield  {title} {\bibinfo {title} {Non-hermitian physics and pt symmetry},\ }\href {https://doi.org/10.1038/nphys4323} {\bibfield  {journal} {\bibinfo  {journal} {Nat. Phys.}\ }\textbf {\bibinfo {volume} {14}},\ \bibinfo {pages} {11} (\bibinfo {year} {2018})}\BibitemShut {NoStop}%
\bibitem [{\citenamefont {Ashida}\ \emph {et~al.}(2020)\citenamefont {Ashida}, \citenamefont {Gong},\ and\ \citenamefont {Ueda}}]{Ashida2020}%
  \BibitemOpen
  \bibfield  {author} {\bibinfo {author} {\bibfnamefont {Y.}~\bibnamefont {Ashida}}, \bibinfo {author} {\bibfnamefont {Z.}~\bibnamefont {Gong}},\ and\ \bibinfo {author} {\bibfnamefont {M.}~\bibnamefont {Ueda}},\ }\bibfield  {title} {\bibinfo {title} {Non-hermitian physics},\ }\href {https://doi.org/10.1080/00018732.2021.1876991} {\bibfield  {journal} {\bibinfo  {journal} {Adv. Phys.}\ }\textbf {\bibinfo {volume} {69}},\ \bibinfo {pages} {249} (\bibinfo {year} {2020})}\BibitemShut {NoStop}%
\bibitem [{\citenamefont {Bergholtz}\ \emph {et~al.}(2021)\citenamefont {Bergholtz}, \citenamefont {Budich},\ and\ \citenamefont {Kunst}}]{Bergholtz2021}%
  \BibitemOpen
  \bibfield  {author} {\bibinfo {author} {\bibfnamefont {E.~J.}\ \bibnamefont {Bergholtz}}, \bibinfo {author} {\bibfnamefont {J.~C.}\ \bibnamefont {Budich}},\ and\ \bibinfo {author} {\bibfnamefont {F.~K.}\ \bibnamefont {Kunst}},\ }\bibfield  {title} {\bibinfo {title} {Exceptional topology of non-hermitian systems},\ }\href {https://doi.org/10.1103/RevModPhys.93.015005} {\bibfield  {journal} {\bibinfo  {journal} {Rev. Mod. Phys.}\ }\textbf {\bibinfo {volume} {93}},\ \bibinfo {pages} {015005} (\bibinfo {year} {2021})}\BibitemShut {NoStop}%
\bibitem [{\citenamefont {Guo}\ \emph {et~al.}(2009)\citenamefont {Guo}, \citenamefont {Salamo}, \citenamefont {Duchesne}, \citenamefont {Morandotti}, \citenamefont {Volatier-Ravat}, \citenamefont {Aimez}, \citenamefont {Siviloglou},\ and\ \citenamefont {Christodoulides}}]{Guo2009}%
  \BibitemOpen
  \bibfield  {author} {\bibinfo {author} {\bibfnamefont {A.}~\bibnamefont {Guo}}, \bibinfo {author} {\bibfnamefont {G.~J.}\ \bibnamefont {Salamo}}, \bibinfo {author} {\bibfnamefont {D.}~\bibnamefont {Duchesne}}, \bibinfo {author} {\bibfnamefont {R.}~\bibnamefont {Morandotti}}, \bibinfo {author} {\bibfnamefont {M.}~\bibnamefont {Volatier-Ravat}}, \bibinfo {author} {\bibfnamefont {V.}~\bibnamefont {Aimez}}, \bibinfo {author} {\bibfnamefont {G.~A.}\ \bibnamefont {Siviloglou}},\ and\ \bibinfo {author} {\bibfnamefont {D.~N.}\ \bibnamefont {Christodoulides}},\ }\bibfield  {title} {\bibinfo {title} {Observation of pt-symmetry breaking in complex optical potentials},\ }\href {https://doi.org/10.1103/PhysRevLett.103.093902} {\bibfield  {journal} {\bibinfo  {journal} {Phys. Rev. Lett.}\ }\textbf {\bibinfo {volume} {103}},\ \bibinfo {pages} {093902} (\bibinfo {year} {2009})}\BibitemShut {NoStop}%
\bibitem [{\citenamefont {Rüter}\ \emph {et~al.}(2010)\citenamefont {Rüter}, \citenamefont {Makris}, \citenamefont {El-Ganainy}, \citenamefont {Christodoulides}, \citenamefont {Segev},\ and\ \citenamefont {Kip}}]{Ruter2010}%
  \BibitemOpen
  \bibfield  {author} {\bibinfo {author} {\bibfnamefont {C.~E.}\ \bibnamefont {Rüter}}, \bibinfo {author} {\bibfnamefont {K.~G.}\ \bibnamefont {Makris}}, \bibinfo {author} {\bibfnamefont {R.}~\bibnamefont {El-Ganainy}}, \bibinfo {author} {\bibfnamefont {D.~N.}\ \bibnamefont {Christodoulides}}, \bibinfo {author} {\bibfnamefont {M.}~\bibnamefont {Segev}},\ and\ \bibinfo {author} {\bibfnamefont {D.}~\bibnamefont {Kip}},\ }\bibfield  {title} {\bibinfo {title} {Observation of parity-time symmetry in optics},\ }\href {https://doi.org/10.1038/nphys1515} {\bibfield  {journal} {\bibinfo  {journal} {Nat. Phys.}\ }\textbf {\bibinfo {volume} {6}},\ \bibinfo {pages} {192} (\bibinfo {year} {2010})}\BibitemShut {NoStop}%
\bibitem [{\citenamefont {Lin}\ \emph {et~al.}(2011)\citenamefont {Lin}, \citenamefont {Ramezani}, \citenamefont {Eichelkraut}, \citenamefont {Kottos}, \citenamefont {Cao},\ and\ \citenamefont {Christodoulides}}]{Lin2011}%
  \BibitemOpen
  \bibfield  {author} {\bibinfo {author} {\bibfnamefont {Z.}~\bibnamefont {Lin}}, \bibinfo {author} {\bibfnamefont {H.}~\bibnamefont {Ramezani}}, \bibinfo {author} {\bibfnamefont {T.}~\bibnamefont {Eichelkraut}}, \bibinfo {author} {\bibfnamefont {T.}~\bibnamefont {Kottos}}, \bibinfo {author} {\bibfnamefont {H.}~\bibnamefont {Cao}},\ and\ \bibinfo {author} {\bibfnamefont {D.~N.}\ \bibnamefont {Christodoulides}},\ }\bibfield  {title} {\bibinfo {title} {Unidirectional invisibility induced by pt-symmetric periodic structures},\ }\href {https://doi.org/10.1103/PhysRevLett.106.213901} {\bibfield  {journal} {\bibinfo  {journal} {Phys. Rev. Lett.}\ }\textbf {\bibinfo {volume} {106}},\ \bibinfo {pages} {213901} (\bibinfo {year} {2011})}\BibitemShut {NoStop}%
\bibitem [{\citenamefont {Liertzer}\ \emph {et~al.}(2012)\citenamefont {Liertzer}, \citenamefont {Ge}, \citenamefont {Cerjan}, \citenamefont {Stone}, \citenamefont {T{\"u}reci},\ and\ \citenamefont {Rotter}}]{Liertzer2012}%
  \BibitemOpen
  \bibfield  {author} {\bibinfo {author} {\bibfnamefont {M.}~\bibnamefont {Liertzer}}, \bibinfo {author} {\bibfnamefont {L.}~\bibnamefont {Ge}}, \bibinfo {author} {\bibfnamefont {A.}~\bibnamefont {Cerjan}}, \bibinfo {author} {\bibfnamefont {A.~D.}\ \bibnamefont {Stone}}, \bibinfo {author} {\bibfnamefont {H.~E.}\ \bibnamefont {T{\"u}reci}},\ and\ \bibinfo {author} {\bibfnamefont {S.}~\bibnamefont {Rotter}},\ }\bibfield  {title} {\bibinfo {title} {Pump-induced exceptional points in lasers},\ }\href {https://doi.org/10.1103/PhysRevLett.108.173901} {\bibfield  {journal} {\bibinfo  {journal} {Phys. Rev. Lett.}\ }\textbf {\bibinfo {volume} {108}},\ \bibinfo {pages} {173901} (\bibinfo {year} {2012})}\BibitemShut {NoStop}%
\bibitem [{\citenamefont {Brandstetter}\ \emph {et~al.}(2014)\citenamefont {Brandstetter}, \citenamefont {Liertzer}, \citenamefont {Deutsch}, \citenamefont {Klang}, \citenamefont {Schöberl}, \citenamefont {T{\"u}reci}, \citenamefont {Strasser}, \citenamefont {Unterrainer},\ and\ \citenamefont {Rotter}}]{Brandstetter2014}%
  \BibitemOpen
  \bibfield  {author} {\bibinfo {author} {\bibfnamefont {M.}~\bibnamefont {Brandstetter}}, \bibinfo {author} {\bibfnamefont {M.}~\bibnamefont {Liertzer}}, \bibinfo {author} {\bibfnamefont {C.}~\bibnamefont {Deutsch}}, \bibinfo {author} {\bibfnamefont {P.}~\bibnamefont {Klang}}, \bibinfo {author} {\bibfnamefont {J.}~\bibnamefont {Schöberl}}, \bibinfo {author} {\bibfnamefont {H.~E.}\ \bibnamefont {T{\"u}reci}}, \bibinfo {author} {\bibfnamefont {G.}~\bibnamefont {Strasser}}, \bibinfo {author} {\bibfnamefont {K.}~\bibnamefont {Unterrainer}},\ and\ \bibinfo {author} {\bibfnamefont {S.}~\bibnamefont {Rotter}},\ }\bibfield  {title} {\bibinfo {title} {Reversing the pump dependence of a laser at an exceptional point},\ }\href {https://doi.org/10.1038/ncomms5034} {\bibfield  {journal} {\bibinfo  {journal} {Nat. Commun.}\ }\textbf {\bibinfo {volume} {5}},\ \bibinfo {pages} {4034} (\bibinfo {year} {2014})}\BibitemShut {NoStop}%
\bibitem [{\citenamefont {Ramezani}\ \emph {et~al.}(2014)\citenamefont {Ramezani}, \citenamefont {Li}, \citenamefont {Wang},\ and\ \citenamefont {Zhang}}]{Ramezani2014}%
  \BibitemOpen
  \bibfield  {author} {\bibinfo {author} {\bibfnamefont {H.}~\bibnamefont {Ramezani}}, \bibinfo {author} {\bibfnamefont {H.-K.}\ \bibnamefont {Li}}, \bibinfo {author} {\bibfnamefont {Y.}~\bibnamefont {Wang}},\ and\ \bibinfo {author} {\bibfnamefont {X.}~\bibnamefont {Zhang}},\ }\bibfield  {title} {\bibinfo {title} {Unidirectional spectral singularities},\ }\href {https://doi.org/10.1103/PhysRevLett.113.263905} {\bibfield  {journal} {\bibinfo  {journal} {Phys. Rev. Lett.}\ }\textbf {\bibinfo {volume} {113}},\ \bibinfo {pages} {263905} (\bibinfo {year} {2014})}\BibitemShut {NoStop}%
\bibitem [{\citenamefont {Hodaei}\ \emph {et~al.}(2017)\citenamefont {Hodaei}, \citenamefont {Hassan}, \citenamefont {Wittek}, \citenamefont {Garcia-Gracia}, \citenamefont {El-Ganainy}, \citenamefont {Christodoulides},\ and\ \citenamefont {Khajavikhan}}]{Hodaei2017}%
  \BibitemOpen
  \bibfield  {author} {\bibinfo {author} {\bibfnamefont {H.}~\bibnamefont {Hodaei}}, \bibinfo {author} {\bibfnamefont {A.~U.}\ \bibnamefont {Hassan}}, \bibinfo {author} {\bibfnamefont {S.}~\bibnamefont {Wittek}}, \bibinfo {author} {\bibfnamefont {H.}~\bibnamefont {Garcia-Gracia}}, \bibinfo {author} {\bibfnamefont {R.}~\bibnamefont {El-Ganainy}}, \bibinfo {author} {\bibfnamefont {D.~N.}\ \bibnamefont {Christodoulides}},\ and\ \bibinfo {author} {\bibfnamefont {M.}~\bibnamefont {Khajavikhan}},\ }\bibfield  {title} {\bibinfo {title} {Enhanced sensitivity at higher-order exceptional points},\ }\href {https://doi.org/10.1038/nature23280} {\bibfield  {journal} {\bibinfo  {journal} {Nature}\ }\textbf {\bibinfo {volume} {548}},\ \bibinfo {pages} {187} (\bibinfo {year} {2017})}\BibitemShut {NoStop}%
\bibitem [{\citenamefont {Chen}\ \emph {et~al.}(2017)\citenamefont {Chen}, \citenamefont {{\"O}zdemir}, \citenamefont {Zhao}, \citenamefont {Wiersig},\ and\ \citenamefont {Yang}}]{Chen2017}%
  \BibitemOpen
  \bibfield  {author} {\bibinfo {author} {\bibfnamefont {W.}~\bibnamefont {Chen}}, \bibinfo {author} {\bibfnamefont {S.~K.}\ \bibnamefont {{\"O}zdemir}}, \bibinfo {author} {\bibfnamefont {G.}~\bibnamefont {Zhao}}, \bibinfo {author} {\bibfnamefont {J.}~\bibnamefont {Wiersig}},\ and\ \bibinfo {author} {\bibfnamefont {L.}~\bibnamefont {Yang}},\ }\bibfield  {title} {\bibinfo {title} {Exceptional points enhance sensing in an optical microcavity},\ }\href {https://doi.org/10.1038/nature23281} {\bibfield  {journal} {\bibinfo  {journal} {Nature}\ }\textbf {\bibinfo {volume} {548}},\ \bibinfo {pages} {192} (\bibinfo {year} {2017})}\BibitemShut {NoStop}%
\bibitem [{\citenamefont {Schomerus}(2010)}]{Schomerus_PRL2010_Quantum}%
  \BibitemOpen
  \bibfield  {author} {\bibinfo {author} {\bibfnamefont {H.}~\bibnamefont {Schomerus}},\ }\bibfield  {title} {\bibinfo {title} {Quantum noise and self-sustained radiation of $\mathcal{P}\mathcal{T}$-symmetric systems},\ }\href {https://doi.org/10.1103/PhysRevLett.104.233601} {\bibfield  {journal} {\bibinfo  {journal} {Phys. Rev. Lett.}\ }\textbf {\bibinfo {volume} {104}},\ \bibinfo {pages} {233601} (\bibinfo {year} {2010})}\BibitemShut {NoStop}%
\bibitem [{\citenamefont {Schomerus}(2013)}]{Schomerus_PTRSA2013_From}%
  \BibitemOpen
  \bibfield  {author} {\bibinfo {author} {\bibfnamefont {H.}~\bibnamefont {Schomerus}},\ }\bibfield  {title} {\bibinfo {title} {From scattering theory to complex wave dynamics in non-hermitian pt-symmetric resonators},\ }\href {https://doi.org/10.1098/rsta.2012.0194} {\bibfield  {journal} {\bibinfo  {journal} {Philosophical Transactions of the Royal Society A: Mathematical, Physical and Engineering Sciences}\ }\textbf {\bibinfo {volume} {371}},\ \bibinfo {pages} {20120194} (\bibinfo {year} {2013})}\BibitemShut {NoStop}%
\bibitem [{\citenamefont {Chong}\ \emph {et~al.}(2010)\citenamefont {Chong}, \citenamefont {Ge}, \citenamefont {Cao},\ and\ \citenamefont {Stone}}]{PhysRevLett.105.053901}%
  \BibitemOpen
  \bibfield  {author} {\bibinfo {author} {\bibfnamefont {Y.~D.}\ \bibnamefont {Chong}}, \bibinfo {author} {\bibfnamefont {L.}~\bibnamefont {Ge}}, \bibinfo {author} {\bibfnamefont {H.}~\bibnamefont {Cao}},\ and\ \bibinfo {author} {\bibfnamefont {A.~D.}\ \bibnamefont {Stone}},\ }\bibfield  {title} {\bibinfo {title} {Coherent perfect absorbers: Time-reversed lasers},\ }\href {https://doi.org/10.1103/PhysRevLett.105.053901} {\bibfield  {journal} {\bibinfo  {journal} {Phys. Rev. Lett.}\ }\textbf {\bibinfo {volume} {105}},\ \bibinfo {pages} {053901} (\bibinfo {year} {2010})}\BibitemShut {NoStop}%
\bibitem [{\citenamefont {Chong}\ \emph {et~al.}(2011)\citenamefont {Chong}, \citenamefont {Ge},\ and\ \citenamefont {Stone}}]{PhysRevLett.106.093902}%
  \BibitemOpen
  \bibfield  {author} {\bibinfo {author} {\bibfnamefont {Y.~D.}\ \bibnamefont {Chong}}, \bibinfo {author} {\bibfnamefont {L.}~\bibnamefont {Ge}},\ and\ \bibinfo {author} {\bibfnamefont {A.~D.}\ \bibnamefont {Stone}},\ }\bibfield  {title} {\bibinfo {title} {$\mathcal{P}\mathcal{T}$-symmetry breaking and laser-absorber modes in optical scattering systems},\ }\href {https://doi.org/10.1103/PhysRevLett.106.093902} {\bibfield  {journal} {\bibinfo  {journal} {Phys. Rev. Lett.}\ }\textbf {\bibinfo {volume} {106}},\ \bibinfo {pages} {093902} (\bibinfo {year} {2011})}\BibitemShut {NoStop}%
\bibitem [{\citenamefont {Makris}\ \emph {et~al.}(2014)\citenamefont {Makris}, \citenamefont {Ge},\ and\ \citenamefont {T{\"u}reci}}]{Makris2014}%
  \BibitemOpen
  \bibfield  {author} {\bibinfo {author} {\bibfnamefont {K.~G.}\ \bibnamefont {Makris}}, \bibinfo {author} {\bibfnamefont {L.}~\bibnamefont {Ge}},\ and\ \bibinfo {author} {\bibfnamefont {H.~E.}\ \bibnamefont {T{\"u}reci}},\ }\bibfield  {title} {\bibinfo {title} {Anomalous transient amplification of waves in non-normal photonic media},\ }\href {https://doi.org/10.1103/PhysRevX.4.041044} {\bibfield  {journal} {\bibinfo  {journal} {Phys. Rev. X}\ }\textbf {\bibinfo {volume} {4}},\ \bibinfo {pages} {041044} (\bibinfo {year} {2014})}\BibitemShut {NoStop}%
\bibitem [{\citenamefont {Wanjura}\ \emph {et~al.}(2020)\citenamefont {Wanjura}, \citenamefont {Brunelli},\ and\ \citenamefont {Nunnenkamp}}]{Nature2020}%
  \BibitemOpen
  \bibfield  {author} {\bibinfo {author} {\bibfnamefont {C.~C.}\ \bibnamefont {Wanjura}}, \bibinfo {author} {\bibfnamefont {M.}~\bibnamefont {Brunelli}},\ and\ \bibinfo {author} {\bibfnamefont {A.}~\bibnamefont {Nunnenkamp}},\ }\bibfield  {title} {\bibinfo {title} {Topological framework for directional amplification in driven-dissipative cavity arrays},\ }\bibfield  {journal} {\bibinfo  {journal} {Nat. Commun.}\ }\textbf {\bibinfo {volume} {11}},\ \href {https://doi.org/10.1038/s41467-020-16863-9} {10.1038/s41467-020-16863-9} (\bibinfo {year} {2020})\BibitemShut {NoStop}%
\bibitem [{\citenamefont {Xue}\ \emph {et~al.}(2021)\citenamefont {Xue}, \citenamefont {Li}, \citenamefont {Hu}, \citenamefont {Song},\ and\ \citenamefont {Wang}}]{Xue2021}%
  \BibitemOpen
  \bibfield  {author} {\bibinfo {author} {\bibfnamefont {W.-T.}\ \bibnamefont {Xue}}, \bibinfo {author} {\bibfnamefont {M.-R.}\ \bibnamefont {Li}}, \bibinfo {author} {\bibfnamefont {Y.-M.}\ \bibnamefont {Hu}}, \bibinfo {author} {\bibfnamefont {F.}~\bibnamefont {Song}},\ and\ \bibinfo {author} {\bibfnamefont {Z.}~\bibnamefont {Wang}},\ }\bibfield  {title} {\bibinfo {title} {Simple formulas of directional amplification from non-bloch band theory},\ }\href {https://doi.org/10.1103/PhysRevB.103.L241408} {\bibfield  {journal} {\bibinfo  {journal} {Phys. Rev. B}\ }\textbf {\bibinfo {volume} {103}},\ \bibinfo {pages} {L241408} (\bibinfo {year} {2021})}\BibitemShut {NoStop}%
\bibitem [{\citenamefont {Makris}(2021)}]{Makris2021}%
  \BibitemOpen
  \bibfield  {author} {\bibinfo {author} {\bibfnamefont {K.~G.}\ \bibnamefont {Makris}},\ }\bibfield  {title} {\bibinfo {title} {Transient growth and dissipative exceptional points},\ }\href {https://doi.org/10.1103/PhysRevE.104.054218} {\bibfield  {journal} {\bibinfo  {journal} {Phys. Rev. E}\ }\textbf {\bibinfo {volume} {104}},\ \bibinfo {pages} {054218} (\bibinfo {year} {2021})}\BibitemShut {NoStop}%
\bibitem [{\citenamefont {Ryu}(2023)}]{PhysRevA.108.052205}%
  \BibitemOpen
  \bibfield  {author} {\bibinfo {author} {\bibfnamefont {J.-W.}\ \bibnamefont {Ryu}},\ }\bibfield  {title} {\bibinfo {title} {Dynamics in non-hermitian systems with nonreciprocal coupling},\ }\href {https://doi.org/10.1103/PhysRevA.108.052205} {\bibfield  {journal} {\bibinfo  {journal} {Phys. Rev. A}\ }\textbf {\bibinfo {volume} {108}},\ \bibinfo {pages} {052205} (\bibinfo {year} {2023})}\BibitemShut {NoStop}%
\bibitem [{\citenamefont {Mostafazadeh}(2010)}]{Mostafazadeh2010}%
  \BibitemOpen
  \bibfield  {author} {\bibinfo {author} {\bibfnamefont {A.}~\bibnamefont {Mostafazadeh}},\ }\bibfield  {title} {\bibinfo {title} {Pseudo-hermitian representation of quantum mechanics},\ }\href {https://doi.org/10.1142/S0219887810004816} {\bibfield  {journal} {\bibinfo  {journal} {Int. J. Geom. Methods Mod. Phys.}\ }\textbf {\bibinfo {volume} {7}},\ \bibinfo {pages} {1191} (\bibinfo {year} {2010})}\BibitemShut {NoStop}%
\bibitem [{\citenamefont {Brody}(2013)}]{Brody2013}%
  \BibitemOpen
  \bibfield  {author} {\bibinfo {author} {\bibfnamefont {D.~C.}\ \bibnamefont {Brody}},\ }\bibfield  {title} {\bibinfo {title} {Biorthogonal quantum mechanics},\ }\href {https://doi.org/10.1088/1751-8113/47/3/035305} {\bibfield  {journal} {\bibinfo  {journal} {J. Phys. A: Math. Theor.}\ }\textbf {\bibinfo {volume} {47}},\ \bibinfo {pages} {035305} (\bibinfo {year} {2013})}\BibitemShut {NoStop}%
\bibitem [{\citenamefont {Gong}\ \emph {et~al.}(2018)\citenamefont {Gong}, \citenamefont {Ashida}, \citenamefont {Kawabata}, \citenamefont {Takasan}, \citenamefont {Higashikawa},\ and\ \citenamefont {Ueda}}]{Gong2018}%
  \BibitemOpen
  \bibfield  {author} {\bibinfo {author} {\bibfnamefont {Z.}~\bibnamefont {Gong}}, \bibinfo {author} {\bibfnamefont {Y.}~\bibnamefont {Ashida}}, \bibinfo {author} {\bibfnamefont {K.}~\bibnamefont {Kawabata}}, \bibinfo {author} {\bibfnamefont {K.}~\bibnamefont {Takasan}}, \bibinfo {author} {\bibfnamefont {S.}~\bibnamefont {Higashikawa}},\ and\ \bibinfo {author} {\bibfnamefont {M.}~\bibnamefont {Ueda}},\ }\bibfield  {title} {\bibinfo {title} {Topological phases of non-hermitian systems},\ }\href {https://doi.org/10.1103/PhysRevX.8.031079} {\bibfield  {journal} {\bibinfo  {journal} {Phys. Rev. X}\ }\textbf {\bibinfo {volume} {8}},\ \bibinfo {pages} {031079} (\bibinfo {year} {2018})}\BibitemShut {NoStop}%
\bibitem [{\citenamefont {Yao}\ and\ \citenamefont {Wang}(2018)}]{Yao2018}%
  \BibitemOpen
  \bibfield  {author} {\bibinfo {author} {\bibfnamefont {S.}~\bibnamefont {Yao}}\ and\ \bibinfo {author} {\bibfnamefont {Z.}~\bibnamefont {Wang}},\ }\bibfield  {title} {\bibinfo {title} {Edge states and topological invariants of non-hermitian systems},\ }\href {https://doi.org/10.1103/PhysRevLett.121.086803} {\bibfield  {journal} {\bibinfo  {journal} {Phys. Rev. Lett.}\ }\textbf {\bibinfo {volume} {121}},\ \bibinfo {pages} {086803} (\bibinfo {year} {2018})}\BibitemShut {NoStop}%
\bibitem [{\citenamefont {Herviou}\ \emph {et~al.}(2019)\citenamefont {Herviou}, \citenamefont {Regnault},\ and\ \citenamefont {Bardarson}}]{Herviou2019}%
  \BibitemOpen
  \bibfield  {author} {\bibinfo {author} {\bibfnamefont {L.}~\bibnamefont {Herviou}}, \bibinfo {author} {\bibfnamefont {N.}~\bibnamefont {Regnault}},\ and\ \bibinfo {author} {\bibfnamefont {J.~H.}\ \bibnamefont {Bardarson}},\ }\bibfield  {title} {\bibinfo {title} {Entanglement spectrum and symmetries in non-hermitian fermionic non-interacting models},\ }\href {https://doi.org/10.21468/SciPostPhys.7.5.069} {\bibfield  {journal} {\bibinfo  {journal} {SciPost Phys.}\ }\textbf {\bibinfo {volume} {7}},\ \bibinfo {pages} {069} (\bibinfo {year} {2019})}\BibitemShut {NoStop}%
\bibitem [{\citenamefont {Kornich}(2023)}]{Kornich2023}%
  \BibitemOpen
  \bibfield  {author} {\bibinfo {author} {\bibfnamefont {V.}~\bibnamefont {Kornich}},\ }\bibfield  {title} {\bibinfo {title} {Current-voltage characteristics of the normal metal--insulator--pt-symmetric non-hermitian superconductor junction as a probe of non-hermitian formalisms},\ }\href {https://doi.org/10.1103/PhysRevLett.131.116001} {\bibfield  {journal} {\bibinfo  {journal} {Phys. Rev. Lett.}\ }\textbf {\bibinfo {volume} {131}},\ \bibinfo {pages} {116001} (\bibinfo {year} {2023})}\BibitemShut {NoStop}%
\bibitem [{\citenamefont {Xu}\ and\ \citenamefont {Jin}(2023)}]{Xu2023}%
  \BibitemOpen
  \bibfield  {author} {\bibinfo {author} {\bibfnamefont {H.~S.}\ \bibnamefont {Xu}}\ and\ \bibinfo {author} {\bibfnamefont {L.}~\bibnamefont {Jin}},\ }\bibfield  {title} {\bibinfo {title} {Pseudo-hermiticity protects the energy-difference conservation in the scattering},\ }\href {https://doi.org/10.1103/PhysRevResearch.5.L042005} {\bibfield  {journal} {\bibinfo  {journal} {Phys. Rev. Res.}\ }\textbf {\bibinfo {volume} {5}},\ \bibinfo {pages} {L042005} (\bibinfo {year} {2023})}\BibitemShut {NoStop}%
\bibitem [{\citenamefont {Beenakker}(1998)}]{Beenakker1998}%
  \BibitemOpen
  \bibfield  {author} {\bibinfo {author} {\bibfnamefont {C.~W.~J.}\ \bibnamefont {Beenakker}},\ }\bibfield  {title} {\bibinfo {title} {Thermal radiation and amplified spontaneous emission from a random medium},\ }\href {https://doi.org/10.1103/PhysRevLett.81.1829} {\bibfield  {journal} {\bibinfo  {journal} {Phys. Rev. Lett.}\ }\textbf {\bibinfo {volume} {81}},\ \bibinfo {pages} {1829} (\bibinfo {year} {1998})}\BibitemShut {NoStop}%
\bibitem [{\citenamefont {Ryu}\ \emph {et~al.}(2017)\citenamefont {Ryu}, \citenamefont {Myoung},\ and\ \citenamefont {Park}}]{JWRyu_2017}%
  \BibitemOpen
  \bibfield  {author} {\bibinfo {author} {\bibfnamefont {J.-W.}\ \bibnamefont {Ryu}}, \bibinfo {author} {\bibfnamefont {N.}~\bibnamefont {Myoung}},\ and\ \bibinfo {author} {\bibfnamefont {H.~C.}\ \bibnamefont {Park}},\ }\bibfield  {title} {\bibinfo {title} {Antiresonance induced by symmetry-broken contacts in quasi-one-dimensional lattices},\ }\href {https://doi.org/10.1103/PhysRevB.96.125421} {\bibfield  {journal} {\bibinfo  {journal} {Phys. Rev. B}\ }\textbf {\bibinfo {volume} {96}},\ \bibinfo {pages} {125421} (\bibinfo {year} {2017})}\BibitemShut {NoStop}%
\bibitem [{\citenamefont {Paasschens}\ \emph {et~al.}(1996)\citenamefont {Paasschens}, \citenamefont {Misirpashaev},\ and\ \citenamefont {Beenakker}}]{PhysRevB.54.11887}%
  \BibitemOpen
  \bibfield  {author} {\bibinfo {author} {\bibfnamefont {J.~C.~J.}\ \bibnamefont {Paasschens}}, \bibinfo {author} {\bibfnamefont {T.~S.}\ \bibnamefont {Misirpashaev}},\ and\ \bibinfo {author} {\bibfnamefont {C.~W.~J.}\ \bibnamefont {Beenakker}},\ }\bibfield  {title} {\bibinfo {title} {Localization of light: Dual symmetry between absorption and amplification},\ }\href {https://doi.org/10.1103/PhysRevB.54.11887} {\bibfield  {journal} {\bibinfo  {journal} {Phys. Rev. B}\ }\textbf {\bibinfo {volume} {54}},\ \bibinfo {pages} {11887} (\bibinfo {year} {1996})}\BibitemShut {NoStop}%
\end{thebibliography}%

\end{document}